\documentstyle[preprint,aps]{revtex}
\begin{document}
\draft
\title{Dynamical response of a one dimensional quantum wire electron
system }
\author{S. Das Sarma and E. H.\ Hwang}
\address{Department of Physics, University of Maryland, College Park,
Maryland  20742-4111} 
\date{\today}
\maketitle
\begin{abstract}
We provide a self-contained 
theoretical analysis of the dynamical response of a one dimensional 
electron system, as confined in a semiconductor quantum wire, within
the random phase approximation. We carry out a detailed comparison
with the corresponding two and three dimensional situations, and
discuss the peculiarities arising in the one dimensional linear
response from the non-existence of low energy single-particle
excitations and from the linear nature of the long wavelength plasmon
mode. We provide a critical discussion of the analytic properties of
the complex dielectric function in the complex frequency plane. We
investigate the zeros of the complex dielectric function, and
calculate the plasmon dispersion, damping, and plasmon spectral weight
in one dimension. We
consider finite temperature and impurity scattering effects on one
dimensional plasmon dispersion and damping. 
\end{abstract}


\section{Introduction}

In an important recent experiment, Go\~{n}i {\it et al.} \cite {goni}
measured the  
wave vector dispersion of the collective charge density excitation 
(``plasmons'') in a one dimensional electron gas confined 
in GaAs quantum wire structures by using the inelastic resonant 
light scattering spectroscopy. The measured one dimensional plasmon 
dispersion agrees remarkably well with the earlier calculated 
random-phase-approximation(RPA) predictions\cite {li1}, which initially was 
considered surprising in view of the widespread belief that one 
dimensional electron systems are singular Tomonaga-Luttinger liquids 
where standard Fermi liquid type electron gas theories (of which RPA 
is a prototypical example) are inapplicable. The quantitative agreement
between RPA predictions and experimental results was later explained 
by the fact\cite{li2} that the calculated RPA plasmon dispersion and the 
Tomonaga-Lutinger theory for the collective charge density excitations 
(Tomonaga-Lutinger bosons\cite{tl}) of the one dimensional electron 
system are equivalent by virtue of the vanishing of vertex 
corrections\cite{dl} in the irreducible polarizability of the one
dimensional electron gas. Subsequent to the original experimental 
discovery of the one dimensional plasmon by Go\~{n}i {\it et al.}
\cite{goni} several  
theoretical papers have appeared in the literature explaining various 
special features of the experimental data\cite{bonitz,sch,hwang,wen}.

The purpose of this paper is to provide a complete, self-contained 
theoretical analysis of the dynamical response of a one dimensional
(1D) electron system based mostly on RPA. 
It is surprising that in spite of
considerable interest in the subject matter such a comprehensive
analysis does not exist in the literature. By providing rather 
complete results for a well-defined simple model and by extensively 
comparing the one dimensional results to the well-studied two and 
three dimensional RPA results, we hope to dispel some 
misconceptions and erroneous claims made in the recent theoretical 
literature and to accentuate the peculiar
aspects of one dimensional plasmons (as against the corresponding
higher dimensional situations) arising from the severe
phase-space restrictions in one dimension. We emphasize that RPA 
takes on particular significance in one dimension because its domain 
of validity for plasmon dispersion is large 
(up to second-order in the wave vector in one dimension 
as against only first order, as implied trivially by the f-sum rule,
in two and three dimensions).
 
The rest of the paper is organized as follows: In Sec. II we provide 
the basic RPA formalism and applicable formulae; in Sec. III we give
the numerical results for plasmon 
dispersion and damping, and in Sec. IV we provide the dynamical structure factor
(which is a direct measure of the light scattering spectral
intensity); in Sec. V we 
go beyond RPA by including some local field corrections in the Hubbard
approximation, and we conclude in Sec. VI providing a summary of our
results. 

\section{Theory}

The elementary excitation spectrum of an electron gas is given by the
dynamical structure
factor $S(q,\omega)$ which is proportional to 
Im$[\epsilon(q,\omega)^{-1}]$
where $\epsilon(q,\omega)$ is the dynamical dielectric function of the
electron system. Within RPA, the dielectric function
$\epsilon(q,\omega)$ is given by 
\begin{equation}
\epsilon(q,\omega)=1-v_c(q)\Pi_0(q,\omega),
\end{equation}
where $v_c(q)$ is the Coulomb interaction (in wave-vector space), and 
$\Pi_0(q,\omega)$ is the leading-order irreducible electron gas 
polarizability function ({\it i.e.} the so-called bare bubble or the 
Lindhard function in the relevant dimension, Fig. 1(a)). 
In an ideal 1D electron system the
Coulomb interaction  $v_c(q)$ in the wave vector space is
logarithmically divergent.
By using a more realistic finite-width 1D model\cite{li1,ds}, the Fourier
transformed 1D Coulomb interaction is given by
\begin{equation}
v(q,y-y')=\frac{2e^2}{\epsilon_b}K_0(q|y-y'|)
\end{equation}
where $\epsilon_b$ is the background lattice dielectric constant, and
$K_0$ is the zeroth-order modified Bessel function of the second kind.
In the finite-width 1D model we use a square well potential with infinite
barriers at $y=-a/2$ to $y=a/2$. The 1D electron gas is along
the $x$-axis and the wire width in the $y$ ($z$) direction is taken to
be $a$ (zero). In typical quantum wires, the wire thickness along the
growth direction is much smaller than that in the lateral direction,
justifying our 1D finite width model. The Coulomb matrix element in this
square well potential becomes
\begin{equation}
v_c(q)=\frac{2e^2}{\epsilon_b}\int^{a/2}_{-a/2}dy\int^{a/2}_{-a/2}
dy'K_0(q|y-y'|)\cos^2(\frac{\pi y}{a})\cos^2(\frac{\pi y'}{a}).
\end{equation}
Its asymptotic forms\cite{hu} are
\begin{equation}
v_c(q) = \left \{ \begin{array}{ll}
\frac{3\pi e^2}{\epsilon_b|qa|},  & \mbox {for $|qa| \rightarrow \infty$}, \\
\frac{2e^2}{\epsilon_b}[K_0(|qa|) \; + \; 1.9726917 \cdot \cdot
\cdot], & \mbox{for $|qa| \rightarrow 0$},
\end{array}
\right.
\end{equation}
where
$a$ is the cross-sectional confinement size of the quantum wire.
The noninteracting
polarizability can be exactly calculated in any dimension, and we show
here the results for the one dimensional case only. (The corresponding 
two\cite{ando} and three\cite{fetter} dimensional results can be
found in the literature.) 
The 1D polarizability function $\Pi_0(q,\omega)$ is
\begin{equation}
\Pi_0(q,\omega)=(2s+1)\int_C\frac{dk}{2\pi}\frac{f_{eq}(k+q) -
f_{eq}(k)}{ E(k+q)-E(k)-\omega},
\end{equation}
where $s=1/2$ is the spin of the carrier, $E(k)=k^2/2m-\mu$ with chemical
potential $\mu$, and
$f_{eq}(k)$  is the Fermi distribution function with $k$ as the 1D
wave vector. (Throughout this paper $\hbar=k_B=1$.)

Plasmon dispersion and its lifetime or damping is
determined by the zeros of the complex dielectric function such that
\begin{equation}
\epsilon(q,z_{p})=0,
\end{equation}
where $z_p$ is the complex frequency ($z_{p}=\omega_{p}-i\alpha$), where
$\omega_p$ and $\alpha$ are taken to be positive functions of the 1D mode wave
vector $q$. (This ensures stability of the system.) The condition
Re$[z_p]=\omega_p(q)$ gives the plasmon dispersion and
Im$[z_p]=\alpha(q)$ gives the damping rate of the plasmon mode. Since $\alpha$
is positive, the analytic continuation of the dielectric function into
the lower frequency half plane is needed. Then the polarizability
for the complex frequency ($z=\omega-i\alpha$) becomes
\begin{eqnarray}
\Pi_0(q,z)&=&\frac{2m}{q} \left[\int^{\infty}_{-\infty}\frac{dk}{2
\pi}\frac{f_{eq}(k+q) -f_{eq}(k)}{k+\frac{q}{2}-\frac{m}{q}z} +
i \left \{ f_{eq}(\frac{m}{q}z+\frac{q}{2})-f_{eq}(\frac{m}{q}z -
\frac{q}{2})\right \} \theta(\alpha) \right]   \nonumber \\
&=&\frac{m}{\pi q}\left[ \ln (\frac{z^2 - \omega_{-}^2}{z^2 -
\omega_{+}^2}) + 2\pi i \theta(z_{+} - \omega)\theta(\omega -
z_{-})\theta(\alpha)\right],
\end{eqnarray}
where
\begin{equation}
\omega_{\pm}=\frac{k_{F}q}{m}\pm E(q),
\end{equation}
and
\begin{equation}
z_{\pm}=\sqrt{(\frac{k_Fq}{m})^2+\alpha^2}\pm E(q).
\end{equation}

Without any damping ($i.e.$ when $\alpha=0$) we find
the dispersion 
relation for an undamped quantum wire\cite{li1} plasmon at $T=0$ to be
\begin{equation}
\omega_p(q)=\left[\frac{A(q)\omega_+^2(q)-\omega_-^2(q)}{A(q) - 1}
\right]^{1/2}
\end{equation} 
where $A(q)=\exp[q\pi/mv_c(q)]$. In the long wave length limit ($q
\rightarrow 0$) this
mode becomes
\begin{equation}
\omega_p(q)=q\left[ v_F^2+\frac{2}{\pi}v_Fv_c(q)\right]^{1/2} +
O(q^3).
\end{equation}
This RPA plasmon dispersion up to second order in $q$ is the
same as the 
dispersion of the collective charge density 
excitation in the exactly solvable Tomonaga-Luttinger 1D
model\cite{tl}.

Since $\omega_p(q)$ (Eq. (10)) is greater than $\omega_+(q)$, which is
the upper boundary of the electron-hole pair continuum for all $q$, we
expect this mode 
not to decay to electron-hole pairs within RPA. In three
dimensions, however, $\omega_+(q)$ becomes larger than the
plasmon energy at a critical wave vector, $q_c \approx \omega_p/v_F$. If
$q>q_c$, $\epsilon(q,\omega)$ has a finite imaginary part and the
plasmon mode enters the pair continuum regime, becoming damped
consequently (Landau damping). The plasmon mode inside
the Landau damping region decays by emitting single pair excitations
which is now 
allowed by energy-momentum conservation.
For small Im$[\epsilon(q,\omega)]$, this damping rate can be calculated
by taking $z=\omega(q)-i\alpha$ in Eq. (1) (with $\omega$ as a real
frequency,  and $\alpha/\omega \ll 1$) \cite{fetter}, to obtain
\begin{equation}
{\rm Re}[\epsilon(q,\omega_p(q))]=1-v_c(q){\rm
Re}[\Pi_0(q,\omega_p(q))] = 0,
\end{equation}
and
\begin{equation}
\alpha(q)=\omega_p(q){\rm Im}[\Pi_0(q,\omega_p(q))]
\left [ \frac{\partial}{\partial \omega} {\rm Re}[\Pi_0(q,\omega)]
|_{\omega_p(q)}\right]^{-1}.
\end{equation}
Eq. (12) determines the dispersion relation $\omega_p(q)$ of the plasmon,
and Eq. (13) gives an explicit formula for the plasmon damping rate.
In 2D, the plasmon mode approaches the pair continuum and meets the
damping region at some critical wave vector, $q_c$. Its difference from
3D is that the plasma mode simply ceases to exist at the upper boundary of
the pair continuum at $q_c$ .
Since the plasmon does not exist inside the pair
continuum, the analytic continuation of the dispersion relation, which is
useful to explain the damping of the 3D mode, is invalid in 2D.
Within RPA the 2D plasmon mode does not decay by single pair emission
as long as it exists. It simply disappears at $q_c$.
A similar phenomenon is even more striking in 1D. Since the plasmon
mode always lies above the upper bound of the 1D pair continuum, the
1D plasmon is undamped within RPA in a pure system. Because of this
fundamentally undamped character of 1D RPA plasmons, there are no
solutions of Eq. (6) with a finite $\alpha$ in 1D. 
In some sense one can define a ``true' collective mode by the
$zeros$ of the real 
part of the dielectric function, Re$[\epsilon(q,\omega)]=0$. Within
RPA there are always
two zeros of Re$[\epsilon(q,\omega)]$ for $all$ (1D, 2D, 3D)
dimensions - one inside 
the pair continuum and the other outside the pair continuum. Thus, one
may be misled into concluding that there are 
two collective modes corresponding to the two solutions of
Re$[\epsilon(q,\omega)]=0$. The damping of the mode existing inside the
pair continuum is, however, very large, making it an experimentally
unobservable and theoretically uninteresting entity. (Within RPA, the
strength of the mode outside the pair continuum is typically a
thousand times greater than that of the inside mode.) There has been
some recent confusion in the literature\cite{bonitz} regarding this
point, and incorrect claims of there being two collective modes
(corresponding to the two solutions of Re$[\epsilon(q,\omega)] = 0$ as
discussed above) in 1D systems have been made.
We emphasize that the two zeros of Re$[\epsilon(q,\omega)]$, one
inside and one outside the electron-hole pair continuum, exist in all
dimensions, and only the zero lying outside the continuum is a true
collective mode carrying any significant spectral weight.
 
The expression for the 1D polarizability $\Pi_0(q,\omega)$ given in
Eq. (7) assumes that the system is pure and free
from any defects or impurities. In general, impurity scattering causes
the electrons to diffuse instead of moving ballistically. Impurity
scattering 
effects are usually introduced by including impurity ladder diagrams
in the polarizability
consistently with self-energy corrections in the electron Green's
function (Fig 1(b) and (c)). With the
impurity self-energy correction
the dressed Green's function becomes (within the Born
approximation), 
\begin{equation}
G(q,\omega)=\frac{1}{\omega-E(q)+\frac{i}{2\tau} {\rm sgn}(\omega)},
\end{equation}
where $\tau$ is the impurity induced elastic scattering time in the
Boltzmann approximation. 
Then the expression for polarizability $\Pi_{\gamma}(q,\omega)$
including impurity scattering is given by
\begin{equation}
\Pi_{\gamma}(q,\omega)=\int\frac{dE}{2\pi i}
\frac{\Pi(E,q,\omega)}{1 - v^2 \Pi(E,q,\omega)},
\end{equation}
where $v$ is the impurity disorder potential which is assumed to be
short-ranged (isotropic) with its strength given by
\begin{equation}
v^2 = \frac{1}{2\pi N_F \tau},
\end{equation}
where $N_F$ is the density of states per spin at the Fermi level, which
is given by $2m/(\pi k_F)$ in 1D.
In this calculation we use a simple random Gaussian white-noise model
for the impurity disorder potential
\begin{equation}
\overline {v(x)v(x')}=v^2\delta(x-x').
\end{equation}
In Eq. (15), the leading order polarizability kernel in the presence of
impurity scattering, $\Pi(E,q,\omega)$, is given by
\begin{equation}
\Pi(E,q,\omega)=\sum_{k} G(k,E)G(k+q,E+\omega).
\end{equation}
In doing the $k$ integral we approximate \cite{aaa}
\begin{equation}
\sum_{k}[\;\;\;] \simeq N_F \int d\epsilon_k [\;\;\;].
\end{equation}
By using this approximation Eq. (18) becomes for $q \ll k_F$, $0 < \omega
\ll E_F$, and in the weak-scattering limit ($E_F \tau \gg 1$)
\begin{equation}
\Pi(E,q,\omega) = 2\pi i N_F \frac{\theta(-E(E+\omega))}{\omega +
i/\tau} \left [ 1 + \frac{ Dq^2/\tau}{(\omega+i/\tau)^2}
+ O(q^4) \right ],
\end{equation}
where $D=v_{F}^2 \tau$ is the one-dimensional diffusion constant, and
the Heaviside $\theta$ function ensures that the two poles are
in different half-planes. Finally, the polarizability
$\Pi_{\gamma}(q,\omega)$ is 
\begin{equation}
\Pi_{\gamma}(q,\omega) = N_F \frac { Dq^2/\tau} {(\omega + i/\tau)
\left [ \omega \; - \; \frac{i}{\tau} 
\frac{Dq^2/\tau}{(\omega+i/\tau)^2} \right ]}.
\end{equation}
This expression  approximately describes the long wavelength and low
frequency electronic density response
for diffusive ($E_F\tau \gg 1$) electrons. For very large values of $\tau$,
Eq. (21) reduces to the correct expression for the polarizability,
$\Pi_0(q,\omega)$,
of a noninteracting 1D electron system,
\begin{equation}
\Pi_0(q,\omega)=N_F \frac{v_F^2 q^2}{\omega^2}, \;\;\;\;\;\; ({\rm
for} \;\;\; q \ll \omega/v_F).
\end{equation}
For finite $\tau$, 
in particular, in the limit $q,\omega \rightarrow 0$,
$\Pi_{\gamma}(q,\omega)$ shows the pole structure characteristic of
the diffusive regime, 
\begin{equation}
\Pi_{\gamma}(q,\omega)=-N_F \frac{Dq^2}{Dq^2+i\omega}.
\end{equation}
Note that Eq. (21) is only a long wavelength and low frequency
approximation to the full polarizability, Eq. (15), which is, in
general, unknown within this diagrammatic approach, and can only be
determined by solving the extremely difficult integral equation
defined by Eq. (15). 
Since the exact expression for $\Pi_{\gamma}(q,\omega)$ within this
diagrammatic approach is complicated,
we use in the numerical calculation a particle-conserving
approximation to Eq. (15) for arbitrary values of $q$ and $\omega$,
given  by Mermin \cite{mermin}. In this relaxation time conserving
approximation\cite{mermin}, with 
$\gamma=1/\tau$, the polarizability is given by
\begin{equation}
\Pi_{\gamma}(q,\omega) = \frac{ (\omega + i \gamma) \Pi_0(q, \omega +
i \gamma)} { \omega + i \gamma [ \Pi_0(q, \omega+i\gamma)/ \Pi_0(q,0)]}.
\end{equation}
This approximate formula gives the same diffusive long wavelength and
low frequency behavior ($i.e.$ Eq. (21)) as one gets from the
diagrammatic approach. We have used this formula ($i.e.$ Eq. (24)) for
our full numerical 
calculation for all $q$ and $\omega$, taking the impurity scattering
induced broadening $\gamma$ as a constant phenomenological
parameter. (Note that the plasmon broadening $\alpha (q)$ is, in
general, determined by, but different from the impurity broadening
$\gamma$.)

We next examine finite temperature effects on 1D  dynamical response.
Unlike 3D metals, where the Fermi energy ($E_F \sim 10^4
K$) is much larger
than experimental temperatures, making the zero
temperature formalism very accurate even at the room temperature,
artificial semiconductor structures of reduced
dimensionality have Fermi energies which are typically comparable to
experimental temperatures
({\it e.g.}, $E_F=4.4meV\approx 50 K$), and therefore finite
temperature effects should be included in the theory. This is because
typically these
systems have much lower effective electron 
density and effective mass, and much larger lattice dielectric
constant than metals. Finite temperature 
effects are included in our theory by calculating 
$\Pi_0(q,\omega)$ at $T\neq0$, using the trick introduced by
Maldague\cite{mal} for a 2D electron gas. Its comparative 1D expression
is given by\cite{hu}
\begin{eqnarray}
\Pi_0(q,\omega;T,\mu) & = & \int_{e^{-\mu/T}}^{1} \frac{dx}{(x+1)^2}
\Pi_0(q,\omega;T=0,\mu+T\ln(x)) \nonumber \\
                      &   & + \int_0^1 \frac{dx}{(x+1)^2}
\Pi_0(q,\omega;T=0,\mu-T\ln(x))
\end{eqnarray}
where $\mu \equiv \mu(n,T)$ is the finite temperature chemical
potential to be obtained from the total 
number of particles $n=\int dE D(E) f(E)$, using the 1D electron density
of states, $D(E)$, and the Fermi distribution function, $f(E)$.

\section{Numerical Results for plasmon dispersion}

In this section we present our numerical results for 1D plasmon
dispersion and damping, comparing with the corresponding 2D and 3D
results as appropriate. 
In Fig. 2 we show the plasmon dispersion and the Landau damping region
as a function of the wave vector for $r_s=1.4$. Here $r_s=(\frac{3}{4\pi
n})^{1/3} \frac{1}{a_0}$, $(\frac{1}{\pi n})^{1/2} \frac{1}{a_0}$,
and $\frac{8}{\pi^2 n}\frac{1}{a_0}$ is the dimensionless electron gas density
parameter\cite{fetter,mahan} for 3D, 2D, and 1D, respectively, with
$n$ and $a_0$ being 
respectively the electron density and the effective Bohr radius.
For a GaAs 1D system, $r_s=1.4$ corresponds to a density of 
$n=0.56 \times 10^6 cm^{-1}$, $\varepsilon_b=12.7$,
the effective mass of $m=0.067m_e$, and a Fermi energy of $E_F=4.4
meV$. We choose the quantum confinement size $a = 100 \AA$ in all
our results. While we have chosen typical values of $n$ and $a$ for
GaAs quantum wires in the extreme 1D limit, our qualitative results
are completely independent of our choice of parameter values.
In the long wavelength limit the plasmon dispersion has different 
behavior in different dimensions. In one and two dimensions the
plasma frequency goes to zero as 
$q \sqrt{|\ln(q)|}$ (1D) and $\sqrt{q}$ (2D) as $q \rightarrow 0$, but
in three dimensions the plasma frequency is finite in the long
wavelength limit.
The area bounded by the
solid lines (the upper one, $\omega_+=q^2+2q$ and the lower one,
$\omega_-=q^2 - 2q$, with $k_F$ as the wave vector unit) indicates
the 1D electron-hole pair continuum region. Unlike 2D
and 3D, where the pair continua have only upper boundary,
$\omega_+$, the 1D pair continuum has a low energy gap between $0$ and
$\omega_-$. This gap arises from severe phase space restrictions
imposed\cite{hu} by 1D energy-momentum conservation, and leads to the
non-existence of low energy single particle excitations in 1D Fermi
systems. The dispersion of the 1D plasmon mode lies above the upper
boundary $\omega_+$ of the 
pair continuum for all $q$. This means that the plasmon mode does not
decay into electron-hole pair excitations and is undamped in a pure 1D
system within RPA.
Fig. 3(a) shows (in the absence of impurity scattering, $\gamma=0$) the
zeros of the Re$[\epsilon(q,\omega)]$ and 
Im$[\epsilon(q,\omega)]$ for
$q=0.3k_F$ in the ($\omega-\alpha$) complex frequency
($z=\omega-i\alpha$) plane using Eq. (7), which is an analytic
continuation into the lower half energy plane.
There are two branches (Fig. 3(a)) of zeros in
Re$[\epsilon(q,\omega)]$ for finite $\alpha$ up to $\alpha_{\rm
max}=0.38$, whereas
Im$[\epsilon(q,\omega)]=0$ along the $\alpha=0$ line and for $0 < \omega <
\omega_-$ and $ \omega_+ < \omega <\infty$. The condition for the
existence of a plasmon
is ${\rm Re}[\epsilon]={\rm Im}[\epsilon]=0$, $i.e.$ $\epsilon(q,\omega) = 0$,
and therefore only the point P ($\alpha=0$ and $\omega=\omega_p(q)$)
in Fig. 3(a) corresponds to a true plasmon mode. This mode (the point
P) is undamped with $\alpha = 0$.
For finite $\alpha$, even if Im$[\epsilon(q,\omega)]$
changes sign (as it does along the dotted lines) it does not pass
through a zero because its change is
discontinuous (that is, $|{\rm Im}\epsilon|>0$). Since the dynamical
structure factor is proportional to Im$[\epsilon]^{-1}$ when
Re$[\epsilon]=0$, we expect very
small spectral weight in the region where Re$[\epsilon]=0$ and
Im$[\epsilon] \neq 0$ (along the solid line in Fig. 3(a)). The
structure factor is broad and incoherent, rather than a
$\delta$-function collective mode peak, along the solid line in
Fig. 3(a) where Re$[\epsilon] = 0$, Im$[\epsilon]=0$ (See section
IV). By investigating the analytic continuation of the
dielectric function into the lower half of the complex energy plane
Bonitz et. al.\cite{bonitz} recently concluded that there is a
collective mode in 1D with $\alpha \neq 0$. We believe, for reasons
explained above, that this conclusion is erroneous and 1D dynamical
response is qualitatively no different from 2D and 3D in this respect
-- this extra ``mode'' along the solid line in Fig. 3(a) with
Re$[\epsilon]=0$, Im$[\epsilon] \neq 0$ is 
overdamped and is not a collective mode in any sense.
The collective
plasmon mode corresponds to the zero of the total dielectric function
(i.e. both Re$[\epsilon]=0$ and Im$[\epsilon]=0$) and has zero damping
($\alpha = 0$) within RPA in 1D, and
there are no additional plasma modes for finite $\alpha$ in a single
component 1D system. 
The experimentally observed additional mode of ref.\onlinecite{goni} is
successfully explained quantitatively by introducing two-subband
occupancy into the 
model as has recently been done in the literature\cite{hwang}. The
additional mode arises from the interaction 
between two different 1D subband plasma modes - one is the in-phase
mode and the other the out-of-phase mode; a strictly one subband model
including analytic 
continuation can not explain the existence of the additional mode.

The inclusion of impurity scattering ($\gamma \neq 0$) causes
collisional damping of the plasmon mode introducing level broadening
in addition to any Landau damping arising from the decay of the
plasmon to electron-hole pair excitations. This corresponds to complex
zeros of 
the complex dielectric function with finite $\alpha$. Fig. 3(b) shows
Re$[\epsilon(q,\omega)]$ = Im$[\epsilon(q,\omega)] =0$ lines in the
complex frequency plane for $q=0.3k_F$ and an
impurity scattering induced broadening $\gamma=0.1E_F$. Compared with
Fig. 3(a) the point P has shifted to finite $\alpha$ in Fig. 3(b), and
consequently the 1D plasmon has finite damping in the presence of
impurity scattering.
In Fig. 4 the plasmon dispersion ($\omega(q)=\omega_p(q)-i\alpha(q)$)
calculated in the presence of finite impurity scattering
(Eq. (24)) is shown for various impurity scattering rates,
$\gamma$. In Fig. 4 the curves with $\omega > 0 $ give the plasma
frequency or the real part
($\omega_p(q)$), and those with $\omega<0$ give the plasma damping or
the imaginary part ($\alpha$) of the complex zero solutions
($\omega(q)$), $i.e.$ $\epsilon(q,\omega(q)) = 0$. From the figure it is
clear that the plasmon is 
overdamped below a critical wave vector $q <
q_c$. The critical wave vector, $q_c$, below which the plasmon does
not exist due to impurity scattering effects, depends on the density and the
scattering rate
\begin{equation}
q_c=\frac{K\gamma}{|\ln(Ka\gamma)|}
\end{equation}
where $K=\sqrt{m\epsilon_0/8ne^2}$.

It is known\cite{gq} that the 2D plasmon is also overdamped in the
long wavelength limit in the presence of impurity scattering.
The plasma dispersion of the 2D system for small
$q$ and $\omega$ in the presence of a finite collisional broadening
$\gamma$ is easily calculated on the basis of Eq. (24) (using the
corresponding 2D noninteracting polarizability $\Pi_0$\cite{ando}):
\begin{equation}
\omega=-\frac{i\gamma}{2}+\frac{1}{2}\sqrt{-\gamma^2 + 8\sqrt{2}r_sq}.
\end{equation}
The critical wave vector, $q_c$, at which $\omega$ becomes completely
imaginary in 2D is
\begin{equation}
q_c=\frac{\gamma^2}{8\sqrt{2}r_s}.
\end{equation}
We show the 2D plasmon dispersion for
different value of collisional broadening in Fig. 5(a). The 2D results
in Fig. 5(a) are qualitatively similar to the 1D results in
Fig. 4. As in 1D, the plasmon damping  $\alpha(q)$
decreases with increasing $\gamma$ in the overdamped region because of
the suppression of the electron diffusion rate for larger $\gamma$. When the
impurity scattering rate $\gamma$ is small the 2D plasmon dispersion
is hardly 
affected by impurity scattering. A quantitative comparison of Fig. 4
and 5(a) show that for a given range of collisional broadening
$\gamma$, the plasmon dispersion is more strongly affected in 1D than
in 2D by impurity scattering effects. In 3D, the plasma frequency has
a long wavelength gap, with the long wavelength plasma frequency being
finite in contrast to 1D and 2D systems. Impurity scattering, therefore,
has no qualitative effect on the 3D plasma dispersion (Fig. 5(b)),
only small quantitative corrections. For small $\gamma$, the 3D
plasmon is not overdamped for small $q$, and, therefore, there is no
long wavelength critical wave vector, $q_c$, in 3D below which the 3D
plasmon vanishes. (On the other hand, as discussed earlier, there is a
Landau damping induced large wave vector cut off, $q_c$, for 3D
plasmons, and the 3D plasmon ceases to be a well defined mode for $q >
q_c$, with $q_c$ being the wave vector at which the 3D plasmon
dispersion enters the electron-hole single particle continuum -- see
Fig. 2, for example.)

Fig. 6 shows the finite temperature effect on the plasmon dispersion.
Unlike impurity scattering effects the plasmon dispersion is only
slightly affected by finite temperature effects. Also, in contrast
to impurity scattering effect where increasing $\gamma$ decreases the
plasmon frequency, increasing temperature actually enhances the 1D
plasmon dispersion.
This upward shift in energy with
increasing temperature is a well-known phenomenon in plasma 
physics\cite{kt}, and happens in all dimensions. Thus, effects of
finite collisional broadening and temperature are qualitatively
different on plasma dispersion. In Fig. 6(b) we show our calculated 1D
plasmon dispersion and damping in the presence of both non-zero
temperature and impurity scattering.

\section{Dynamical Structure Factor}

The dielectric function $\epsilon(q,\omega)$ is
related\cite{fetter,mahan} to the 
dynamical structure factor $S(q,\omega)$ by
\begin{equation}
S(q,\omega)=-\frac{1}{n v_c(q)}{\rm Im} \left [ \frac{1}
{\epsilon(q,\omega)} \right ],
\end{equation}
where $n$ is the 1D density and $v_c$ is the Coulomb
interaction, as in Eq. (1). 
$S(q,\omega)$ is, therefore, the spectral weight of the dielectric
function.  
The dynamical structure factor gives a direct measure of the density
fluctuation spectrum of the electron gas.
Thus, many experiments such as inelastic electron 
and Raman scattering spectroscopies directly measure
$S(q,\omega)$. Equivalently, $S(q,\omega)$ is a measure of the
spectral strength of various elementary excitations of the electron
gas. An undamped plasmon shows up as a $\delta$-function peak in
$S(q,\omega)$ indicating the existence of a simple zero of
$\epsilon(q,\omega)$. The damped plasmon ($\alpha \neq 0$), however,
corresponds at best to a broadened peak in $S(q,\omega)$ -- for larger
broadening, as for $q < q_c$ in 1D and 2D with impurity scattering,
the plasmon is overdamped and there is no peak in $S(q,\omega)$.
Note that $S(q,\omega) \propto {\rm Im}[\epsilon(q,\omega)^{-1}]$, and
for our purpose we mostly show results for
Im$[\epsilon(q,\omega)^{-1}]$ because $v_c(q)$ in 1D depends on the
details of the electron confinement model (cf. Eq. (3)).

In Fig. 7 the calculated 1D RPA Im$[\epsilon(q,\omega)^{-1}]$ is
plotted for different wave vectors
as a function of the frequency
for a clean system at $T=0$. When both Re$[\epsilon]$ and
Im$[\epsilon]$ become zero (i.e. $\epsilon(q,\omega) = 0$, which
defines the plasmon mode),
the imaginary part of the inverse dielectric function
Im$[\epsilon(q,\omega)^{-1}]$ is a
$\delta$-function with the strength 
\begin{equation}
W(q)=\frac{\pi} {|\partial {\rm Re}[\epsilon(q,\omega)] /
\partial \omega|_{\omega=\omega_p(q)}},
\end{equation}
where $\omega_p(q)$ is the plasmon dispersion for a
clean electron gas. The sharp $\delta$-function like peak in Fig. 7
corresponds to the well-defined 1D plasmon mode. This figure also shows
the electron-hole single particle excitation (SPE) continua, which show
up as weak broad (incoherent) structures in $S(q,\omega)$.
The continua in Fig. 7 have been enhanced by a factor of 10 to make
them observable on the scale of the plasmons. Note that there is only
one observed peak, the plasmon mode, in Fig. 7 with any appreciable
spectral weight. In contrast to 2D and 3D systems, this remains true
in 1D systems up to rather high wave vectors ($\geq 2k_F$) because of
the severe suppression of SPE. This suppression of SPE continua (up to
rather high wave vectors) is a specific characteristic of the 1D electron
system. (Note that there is only a single plasmon peak in Fig. 7 in
contrast to the erroneous claims of ref. \onlinecite{bonitz}.) The
{\it low}
wave vector suppression of SPE is a generic feature of dielectric
response because it follows quite generally from the $f$-sum
rule\cite{mahan} which is a direct consequence of particle
conservation. At long wavelengths ($q \rightarrow 0$), the plasmon
exhausts the $f$-sum rule, and therefore, carries
all the spectral weight. In higher dimensions, SPE continua start
gaining spectral weights as $q$ increases and eventually ($q \geq
k_F$) become the dominant elementary excitations. In 1D by contrast,
the plasmon is the dominant excitation up to rather high wave vectors
($q \geq 2k_F$) and the SPE continua are strongly suppressed. For more
on $S(q,\omega)$ in 3D \cite{mahan} and 2D \cite{jain} we refer to the
existing literature\cite{mahan,jain}.

Fig. 8 shows the calculated weight (Eq. (30)) of the plasmon mode in
different dimensions as a function of $q$. In 1D (Fig. 8(a)) as
$q\rightarrow 0$ 
$W(q)$ approaches zero because the plasma frequency vanishes. It has a
maximum strength around
$q=2k_F$ for $r_s=1.4$. For $q \rightarrow
\infty$ $W(q)$ approaches zero slowly. 
Note that this absolute vanishing of $W(q)$ for $q \rightarrow 0$ is only a
manifestation of the vanishing of the 1D plasma frequency at long
wavelength -- the plasmon carries all the spectral weight as $q
\rightarrow 0$ because the plasmon frequency vanishes in the long
wavelength limit. In 3D the long wavelength plasmon has a gap, and
therefore the plasmon weight is finite in 3D even for $q=0$. Using
Eq. (30), the plasmon weight $W(q)$ can be analytically calculated as
$q \rightarrow 0$, with the result: $W(q) = \pi \omega_p(q)/2$ in the
relevant dimension.
Note that the small weight of the plasmon mode at long wavelengths
makes it particularly susceptible to collisional damping effects in 1D
and 2D, as we have discussed before.
 
In Figs. 9 to 12 we show Re$[\epsilon(q,\omega)]$,
Im$[\epsilon(q,\omega)]$, and 
Im$[\epsilon(q,\omega)^{-1}]$ as a function of $\omega$ for
$q=0.2k_F$. Figs. 9 and 10 show the results for $T=0$ (Fig. 9) and
$T=0.5E_F$ (Fig. 10) with various
impurity scattering rates  $\gamma$, and  Figs. 11 and 12 show results
for various
temperatures without any impurity scattering (Fig. 11) and with an
impurity scattering 
rate $\gamma=0.1E_F$ (Fig. 12). Both impurity 
scattering and thermal excitation soften the singularity in the
dynamical dielectric function at 
$\omega_+ = q^2 + 2q$ and $\omega_-=q^2 - 2q$ (with $q$ measured in
$k_F$). This
causes broadening of the plasmon $\delta$-function peak in the
dynamical structure factor. As the
impurity scattering rate $\gamma$ increases the plasmon broadening
increases also. Finite temperature effects on plasmon broadening are
in general small in comparison
with impurity scattering effects. Even a small impurity scattering
($\gamma \approx 0.1E_F$) is sufficient to cause observable plasmon
broadening, but observable thermal broadening of plasmons in the clean
($\gamma = 0$) system
requires a temperature of at least $T=0.5E_F$.

\section{Hubbard Approximation}

To improve upon RPA, which is what we have so far employed in this
paper , and which in general is exact only in the long
wavelength limit, Hubbard\cite{mahan,hubb,jonson} introduced a simple local
field correction $G(q)$ to the RPA by approximately summing the ladder
exchange diagrams instead of using only the noninteracting
polarizability $\Pi_0$ in the dynamical screening:
\begin{equation}
\epsilon_H(q,\omega)=1 - \frac{v_c(q)\Pi_0(q,\omega)}{1 + v_c(q) G(q)
\Pi_0(q,\omega)},
\end{equation}
where
\begin{equation}
G(q) = \frac{1}{2} \frac{v_c(\sqrt{q^2+k_F^2})}{v_c(q)},
\end{equation}
where $v_c$ is the Coulomb interaction in the relevant dimension.
In the limit $q \rightarrow 0$, $G(q) \rightarrow 0$, restoring RPA,
which is exact in this limit. Thus in
the long wavelength limit the local field correction vanishes 
and Eq. (31) becomes the RPA dielectric function of Eq. (1). 
The simple Hubbard local field corrections in 3D, 2D, and 1D can be
written\cite{mahan,hubb,jonson} on the basis of Eq. (32) as
\begin{equation}
\begin{array}{lll}
G_{\rm 2D}(q)& = &\frac{1}{2} \frac{q^2}{q^2+k_F^2}, \\
G_{\rm 3D}(q)& = &\frac{1}{2} \frac{q}{\sqrt{q^2+k_F^2}}, 
\end{array}
\end{equation}
and in 1D
\begin{equation}
G_{\rm 1D}(q) = \frac{1}{2} \frac{v_c(\sqrt{q^2+k_F^2})}{v_c(q)},
\end{equation}
where $v_c(q)$ is now the 1D Coulomb interaction in the quantum wire
as defined by Eq. (3).
In Fig. 13 we show the collective plasmon dispersion as calculated by
using the Hubbard 
approximation (dashed line) and compared with RPA results (solid
line). Dotted 
lines in Fig. 13 indicate the boundaries of the electron-hole pair
continuum. In the long 
wavelength limit the two approximations match exactly, but as $q$
increases the Hubbard plasmon dispersion lies below the RPA plasmon in
all dimensions.
Thus the Hubbard local field correction suppresses the plasmon
frequency at finite $q$ in all dimensions. 
The Hubbard correction, in fact, is larger in lower
dimensions. Whether the Hubbard approximation is an improvement over
RPA in determining the plasmon dispersion is unclear because the
Hubbard local field correction is purely static whereas the exact
local field correction is obviously dynamic, and plasmon dispersion is
manifestly a dynamical property. We have provided here the plasmon Hubbard
approximation results only for the sake of completeness. We
believe that the Hubbard approximation is, in fact, worse than RPA in
determining the plasmon dispersion, at least in 1D and 2D\cite{jonson}.
Within the Tomonaga-Luttinger model\cite{tl} (i.e. linear bare electron energy
dispersion and short-range model interaction), RPA is
known\cite{li2} to be exact because all vertex corrections to the
irreducible polarizability vanish\cite{dl}. For the realistic
Coulomb interaction and the quadratic electron energy dispersion,
which is our 1D electron gas model in this paper, vertex corrections are
not expected to vanish exactly, but should be small. Thus, Hubbard
approximation, which is a crude estimate of ladder vertex corrections,
should not be much of an improvement over RPA in 1D. Comparison with the
experimental results of ref. \onlinecite{goni} indicates that the
experimental 1D plasmon dispersion lies above the RPA curve whereas we
find (Fig. 13) that the Hubbard results lie below the RPA results, and
therefore inclusion of local field corrections does not by any means
improve agreement between theory and experiment. In fact, essentially
exact quantitative agreement between the RPA theory and
experiment\cite{goni} has recently been obtained\cite{hwang} by
incorporating a small occupancy of a higher 1D subband in the
calculation. 

\section{Conclusion}

We have discussed in this article in considerable details the RPA
dynamical response of a one dimensional electron gas as confined, for
example, in an ultra-narrow semiconductor quantum wire structure in
the one subband occupancy limit. Special features of our work are
inclusion of finite impurity scattering and finite temperature effects
in the theory and a fairly comprehensive discussion of
analytic continuation properties of the complex dielectric function in
the complex frequency plane (necessitated by some confusion on this
topic in the recent literature), as well as a detailed comparison between 1D
plasmon properties with corresponding 2D and 3D results. While the
motivation for our work has been the experimental
observation\cite{goni} of 1D plasmons in GaAs quantum wires via the
inelastic light scattering spectroscopy, we have not attempted in this
paper any
comparison between theory and experiment because such a comparison
already exists in the literature\cite{goni,hwang}. It has been
established\cite{hwang} that a complete quantitative and qualitative
understanding of all aspects of the experimental
observations\cite{goni} requires a two-subband model with the second
subband occupied slightly by electrons. In the current paper we
consider the one-subband purely one-dimensional limit which, in fact,
explains the basic features of the experimental observations\cite{goni}
very well, as was already noted in ref. \onlinecite{goni}. Our goal in
this paper is to bring the understanding of 1D dynamical response to
the same level as that already existing for 2D\cite{ando,jain} and 3D
\cite{fetter,mahan}systems.

The two most significant features of 1D dynamical response are (1) the
linear (with small logarithmic corrections depending on the details of the
confinement potential) wave vector dispersion of the long wavelength
collective charge density excitation plasmon mode, and (2) the total
suppression of the electron-hole single particle excitation continua
at low energy (which arises from the severe 1D phase restriction due
to energy-momentum conservation). The 1D plasmon, in fact, exhausts
the $f$-sum rule up to second order in $q/k_F$, and, therefore, the 1D RPA
plasmon dispersion is exact not only in the long wavelength
limit (as it is trivially in all dimensions) but up to large values of
wave vector
($q/k_F \sim 1$). This has earlier been pointed out in the
literature\cite{li2} in a somewhat different context. Our detailed RPA
treatment of 1D dynamical response is thus of considerable
quantitative validity because RPA turns out to
be a more accurate approximation in 1D than in 2D and 3D systems.

Finally, we make some remarks on how the various characteristic
peculiarities of 1D systems, which we neglect uncritically in our
work, are expected to affect our RPA theory of dynamical response. The
three well-known aspects of one dimensionality are Tomonaga-Luttinger
behavior\cite{tl,dl}, Peierls transition and the associated charge
density wave instability\cite{pe}, and Anderson
localization\cite{an}. As mentioned earlier in this article and as has
been emphasized in ref. \onlinecite{li2}, a direct implication of the
1D Tomonaga-Luttinger behavior is that only the bubble diagrams
survive\cite{dl} in the irreducible polarizability, making RPA
essentially an exact approximation for the charge density excitation
mode. Thus, what we refer to as plasmons throughout this article are
the same\cite{li2} as the Tomonaga bosons or the charge collective
excitation modes\cite{tl} of a Tomonaga-Luttinger liquid. Much has been written
on the Peierls transition and associated charge density wave
instability in 1D systems. We just point out that for GaAs quantum
wire based on 1D electron systems, Peierls transition is not a
relevant issue as 
has been emphasized recently in the literature\cite{jr}. This is true
even in the absence of impurity scattering -- inclusion of impurity
scattering only makes Peierls transition and charge density wave
instability issue even more irrelevant\cite{ds} for one dimensional
quantum wire electron systems. In 1D, any disorder, in principle,
completely localizes\cite{an} a noninteracting electron system,
producing exponentially Anderson localized states whereas we have, of
course, assumed delocalized plane wave 1D electron states in this
paper. Again, this 
is not a particularly important issue for our work because in high
mobility GaAs -- based quantum wire systems the calculated
localization lengths\cite{liu} are very long ($ > 10 -100 \mu m$), and
the system behaves for all practical purpose as a delocalized
system. Thus the characteristic peculiarities of 1D systems, which we
have ignored in our work, do not in any significant way affect
our results or conclusions on the dynamical response of quantum wire
based 1D electron systems.

\section*{ACKNOWLEDGMENTS}
This work is supported by the U.S. ARO, the U.S. ONR, and the
NSF-DMR-MRG.

\begin{figure}
\caption{(a) The bare bubble diagram, (b) the dressed Green's
function, and (c) a typical ladder diagram due to impurity scattering.}
\end{figure} 

\begin{figure}
\caption{ Plasma dispersion in different dimensions. Here the
density parameter $r_s=1.4$ is used for all dimensions. The single
particle electron-hole continuum excitation regime in 1D is indicated by
the solid curves.}
\end{figure}

\begin{figure}
\caption{ The zeros of Re$[\epsilon(q,\omega-i\alpha)]$ and
Im$[\epsilon(q,\omega-i\alpha)]$ in the complex $\omega - \alpha$
plane, where $\alpha$ indicates the damping 
of the plasmon mode, for $q=0.3k_F$. Point P corresponds to the
location of the plasmon: ${\rm Re}[\epsilon] = 
{\rm Im}[\epsilon] = 0$. (a) Clean system with the impurity scattering
rate $\gamma = 0$. 
(b) Dirty system with impurity scattering rate $\gamma=0.1E_F$. Note
that P lies on the $\alpha = 0$ line in (a), but is shifted to finite
$\alpha$ in (b).}
\end{figure}

\begin{figure}
\caption{1D plasmon dispersion is shown for various impurity
scattering rates $\gamma$ as given in the figure. Here $\omega >0$
gives the dispersion of the mode and 
$\omega <  0$ the damping of the mode. The overdamping of the
plasmon for $q<q_c$ is clearly seen. Parameters are
$a=100\AA$ and $r_s=1.4$ which corresponds to the 1D density $n=0.56\times
10^6 cm^{-1}$ for GaAs.}
\end{figure}

\begin{figure}
\caption{Plasmon dispersion and damping in (a) 2D
and (b) 3D ($r_s = 1.4$ in both).}
\end{figure}

\begin{figure}
\caption{ (a) 1D plasmon dispersion and damping
without any impurity scattering for various temperatures as shown.
(b) 1D plasmon dispersion and damping at a 
finite temperature $T=0.5E_F$ and for various impurity scattering
rates $\gamma$ as shown ($r_s = 1.4$).}
\end{figure}

\begin{figure}
\caption{Im$[\epsilon(q,\omega)^{-1}]$ in 1D as a function
of frequency $\omega$ for different wave vectors $q$. The $\delta$-function
peaks are the plasmon modes with their spectral weights written above
the peaks. 
The strength of the electron-hole continuum (the broad incoherent bump
in each figure) has been enhanced by a factor ten for visual clarity.}
\end{figure}

\begin{figure}
\caption{Calculated weights $W(q)$ of the plasmon mode in different
dimensions: (a) 1D, (b) 2D, and (c) 3D.}
\end{figure}

\begin{figure}
\caption{Shows for a fixed wave vector $q=0.2k_F$: (a)
Re$[\epsilon(q,\omega)]$; (b) 
Im$[\epsilon(q,\omega)]$; and 
(c) Im$[\epsilon(q,\omega)^{-1}]$ as a function of frequency at $T=0$
with  various impurity scattering rates $\gamma$. 
In (c) the electron-hole pair continuum for $\gamma=0$ (solid line)
has been
enhanced by a factor 10 for the sake of clarity. The vertical arrow in
(c) denotes the plasmon $\delta$-function peak.} 
\end{figure}

\begin{figure}
\caption{Shows for a fixed wave vector $q=0.2k_F$ the calculated (a)
Re$[\epsilon(q,\omega)]$; (b) Im$[\epsilon(q,\omega)]$; and
(c) Im$[\epsilon(q,\omega)^{-1}]$ as a function of frequency
$\omega$ at a finite
temperature $T=0.5E_F$ with  various impurity scattering rates
$\gamma$ as shown.}
\end{figure}

\begin{figure}
\caption{Shows for a fixed wave vector $q=0.2k_F$ the calculated (a)
Re$[\epsilon(q,\omega)]$; (b) Im$[\epsilon(q,\omega)]$; and
(c) Im$[\epsilon(q,\omega)^{-1}]$ as a function of frequency
$\omega$ without
any impurity scattering ($\gamma=0$) for various temperatures as shown. }
\end{figure}

\begin{figure}
\caption{Shows for a wave vector $q=0.2k_F$ the calculated (a)
Re$[\epsilon(q,\omega)]$; (b) Im$[\epsilon(q,\omega)]$; and
(c) Im$[\epsilon(q,\omega)^{-1}]$ as a function of frequency
$\omega$ with
an impurity scattering rate $\gamma=0.1E_F$ for various
temperatures as shown.}
\end{figure}

\begin{figure}
\caption{Collective plasmon dispersion within the Hubbard
approximation (dashed lines) and within RPA (solid lines) for 
different dimensions: (a) 1D; (b) 2D; and (c) 3D. 
Dotted lines indicate the boundaries of the electron-hole pair continua.} 
\end{figure}
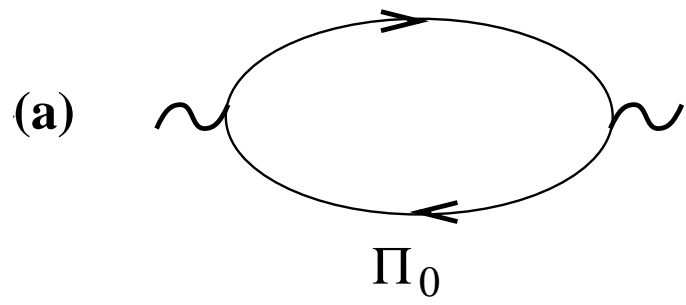
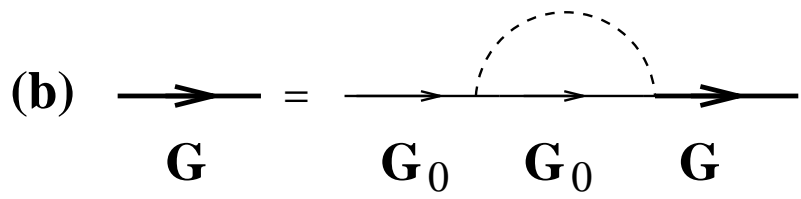
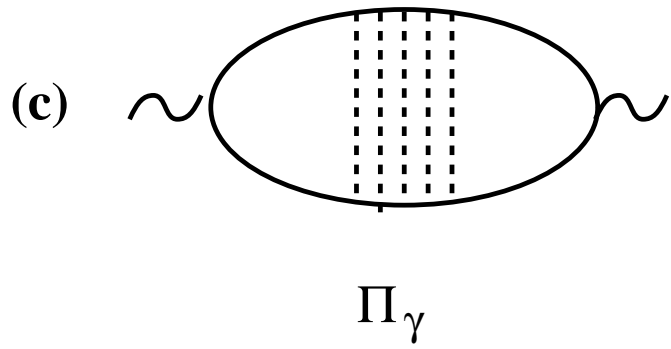
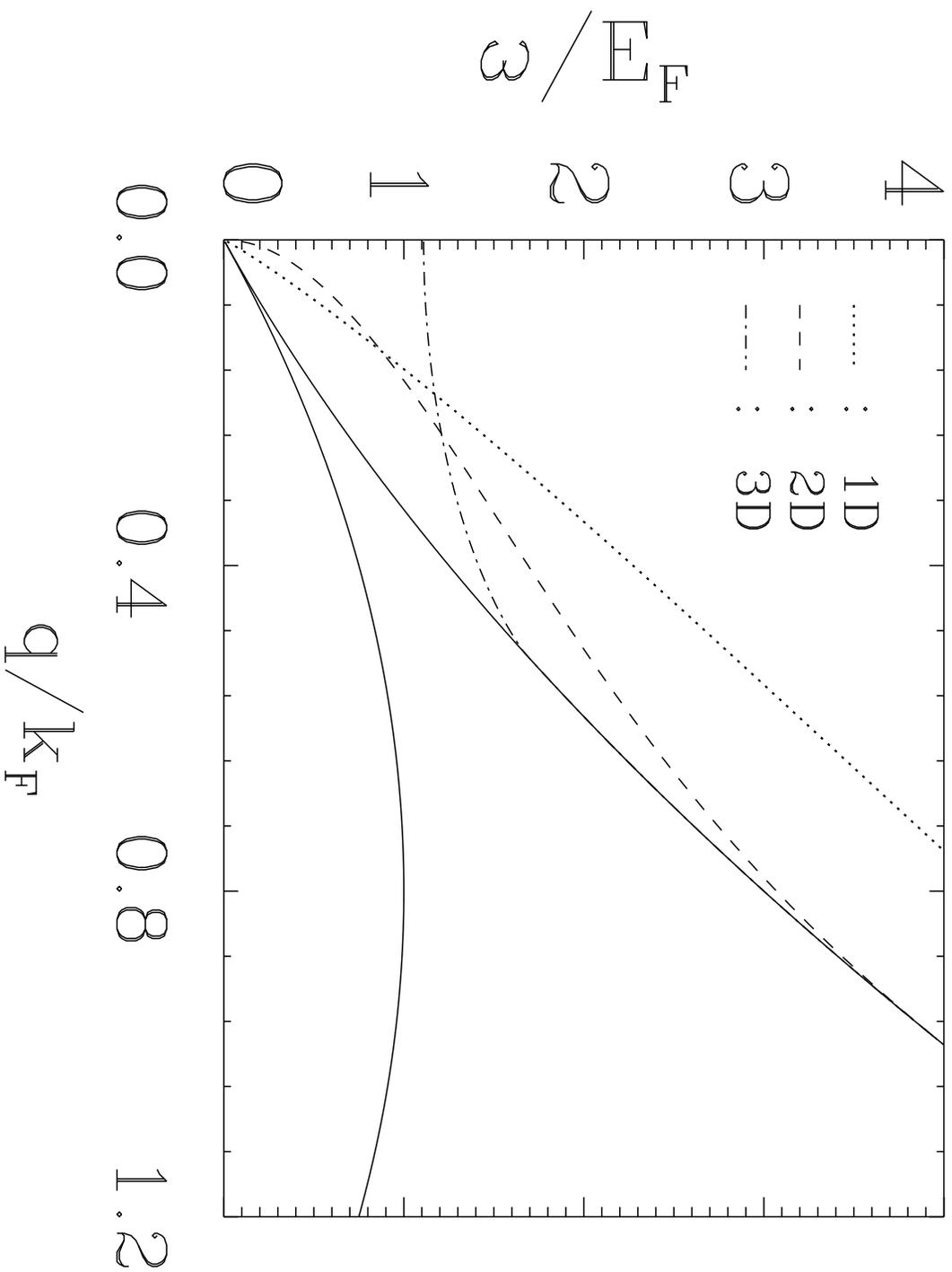
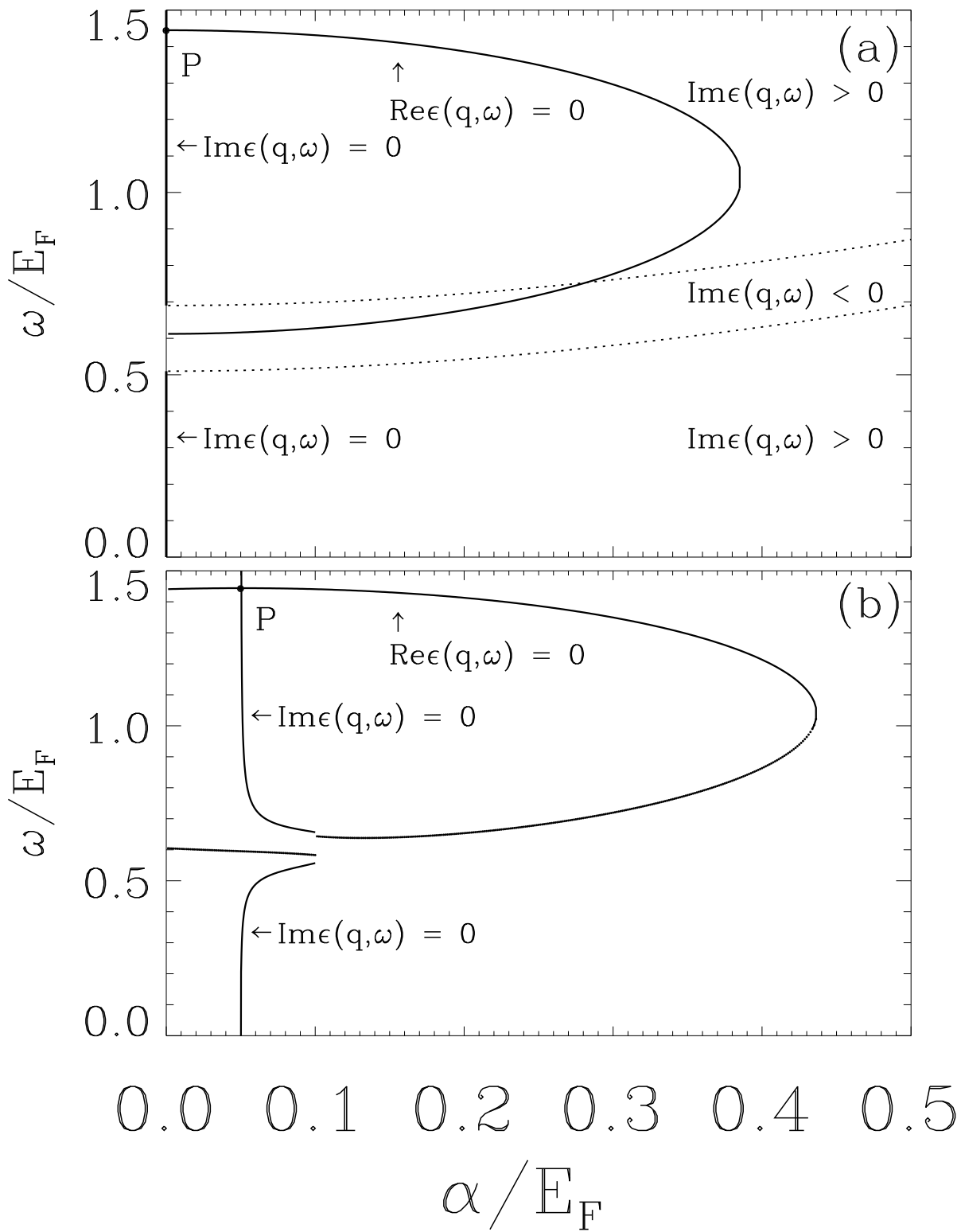
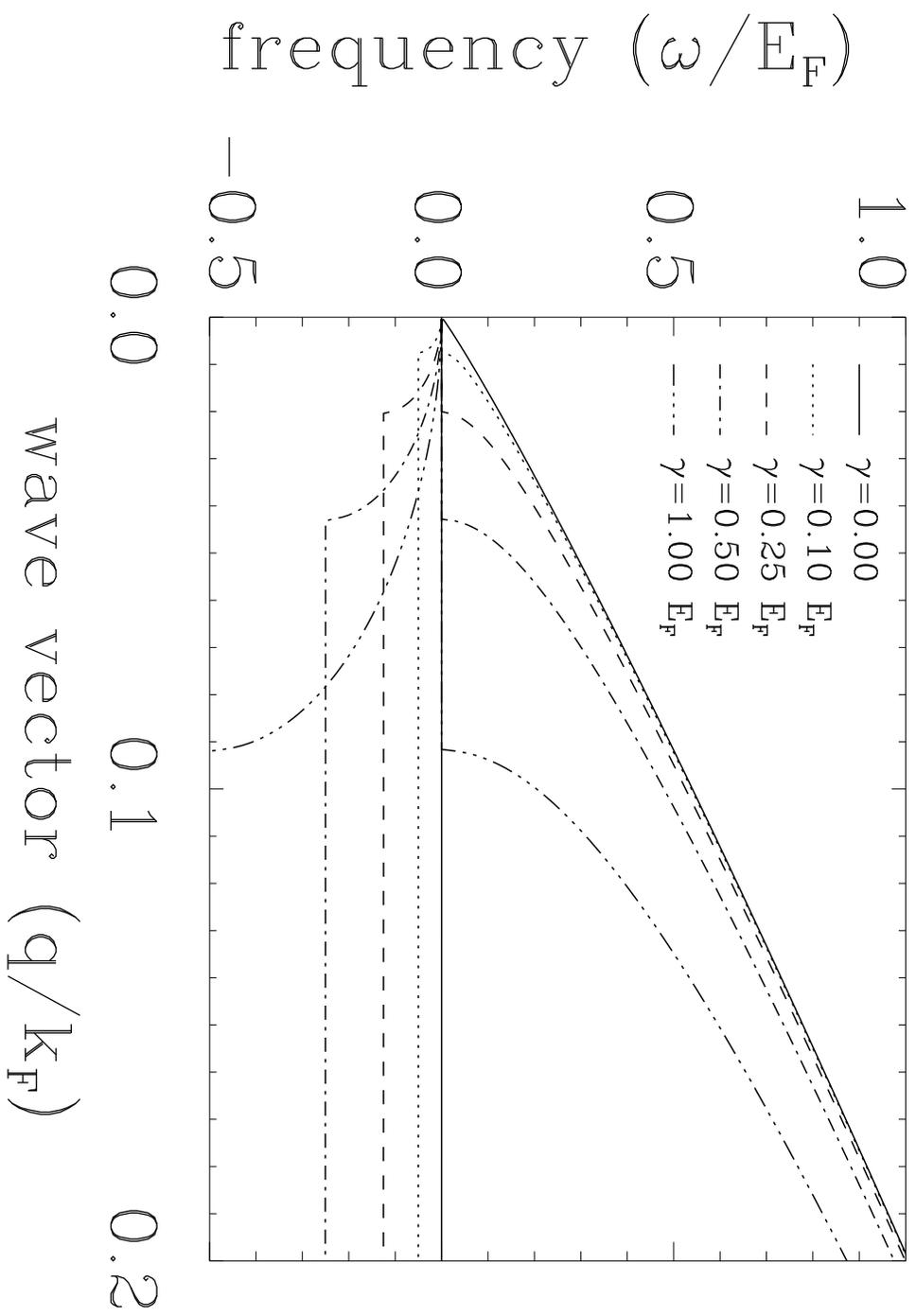
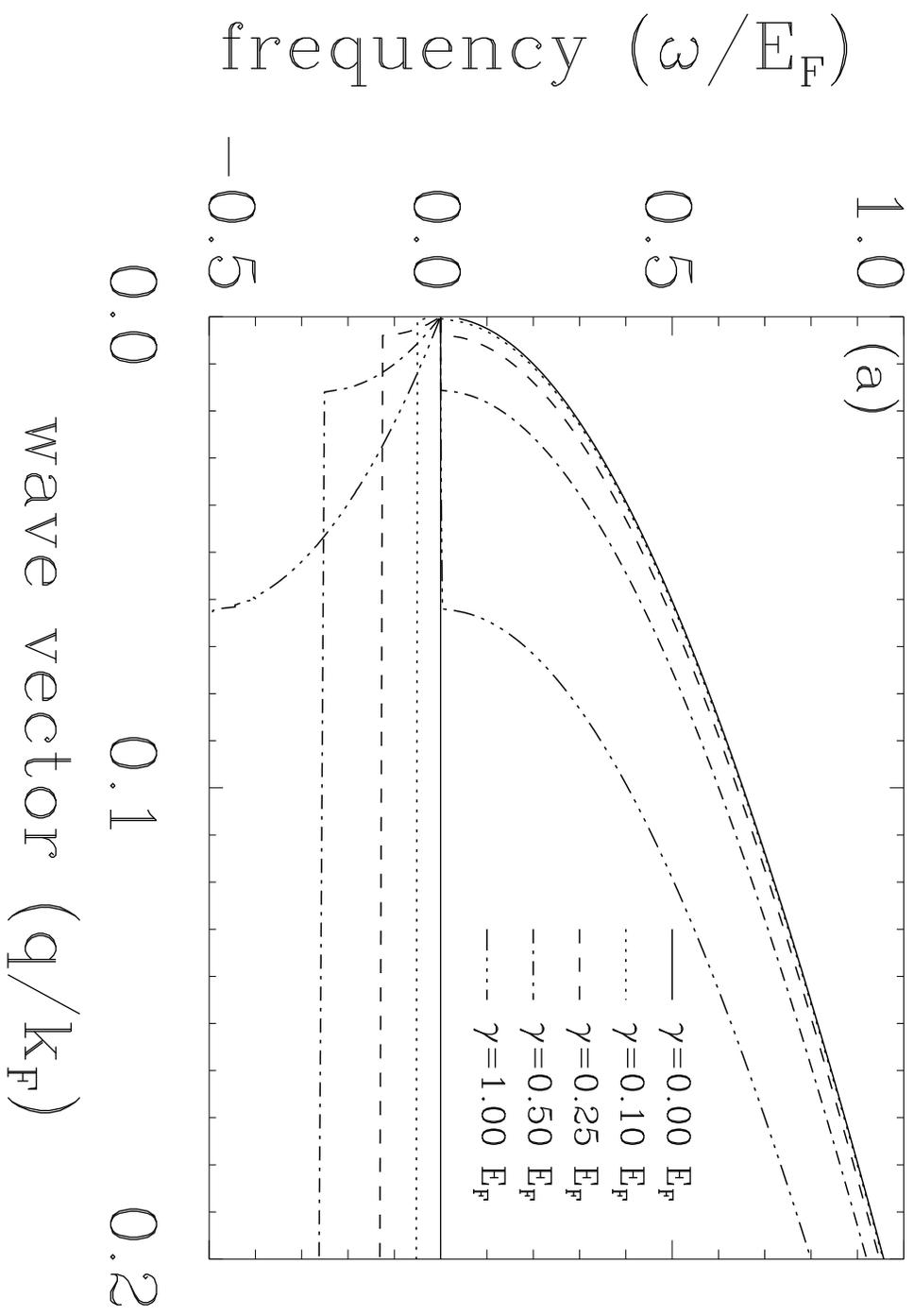
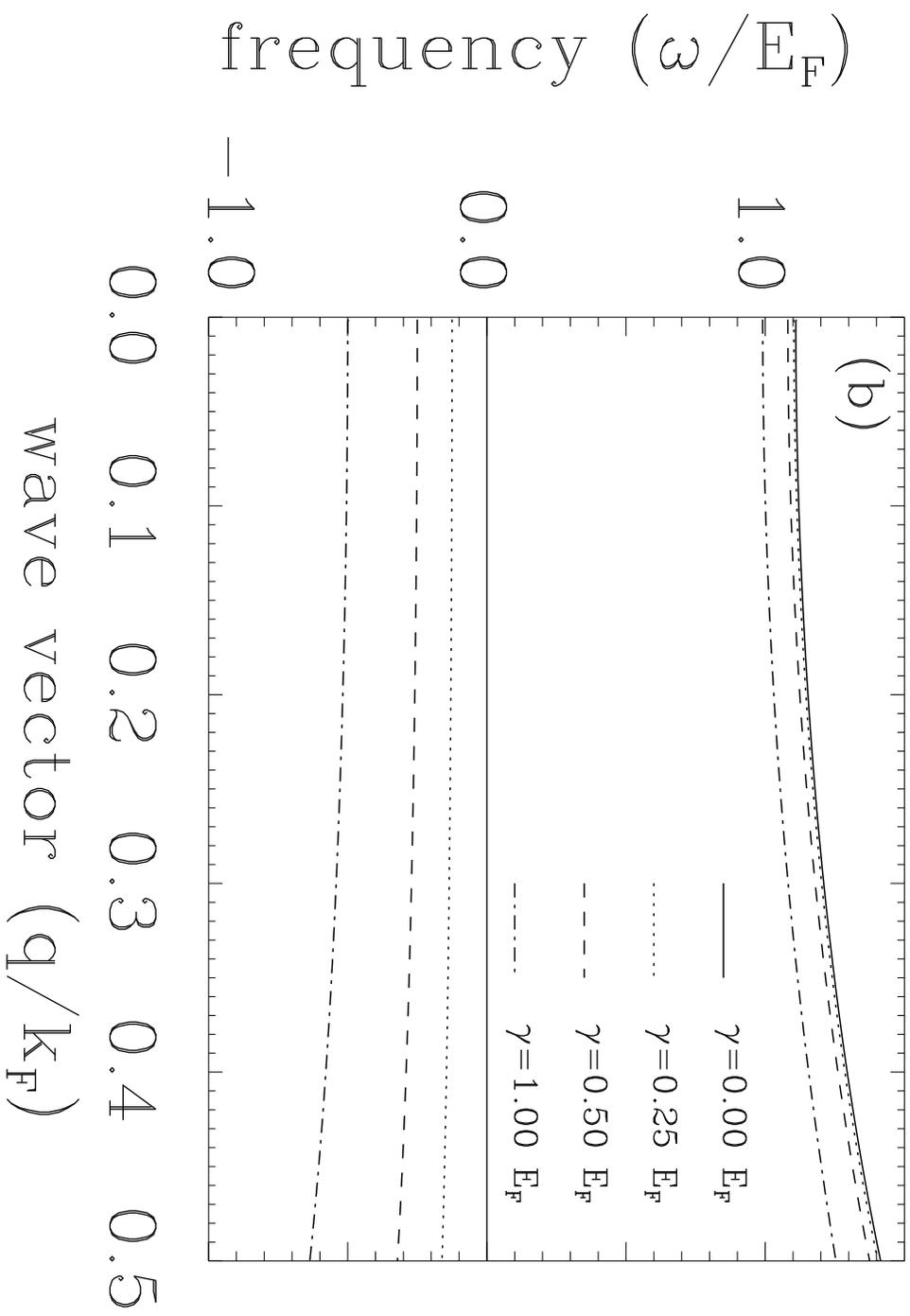
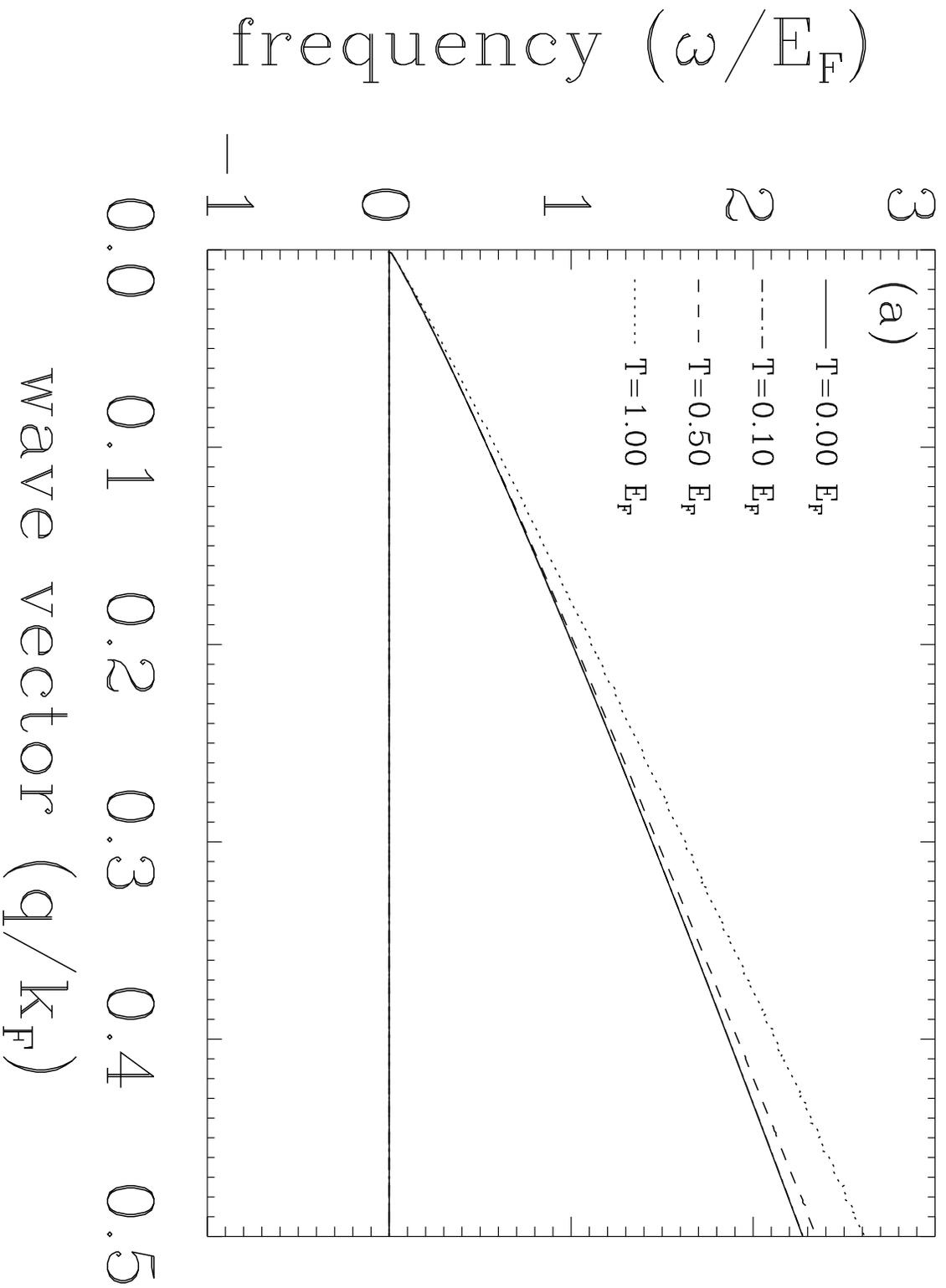
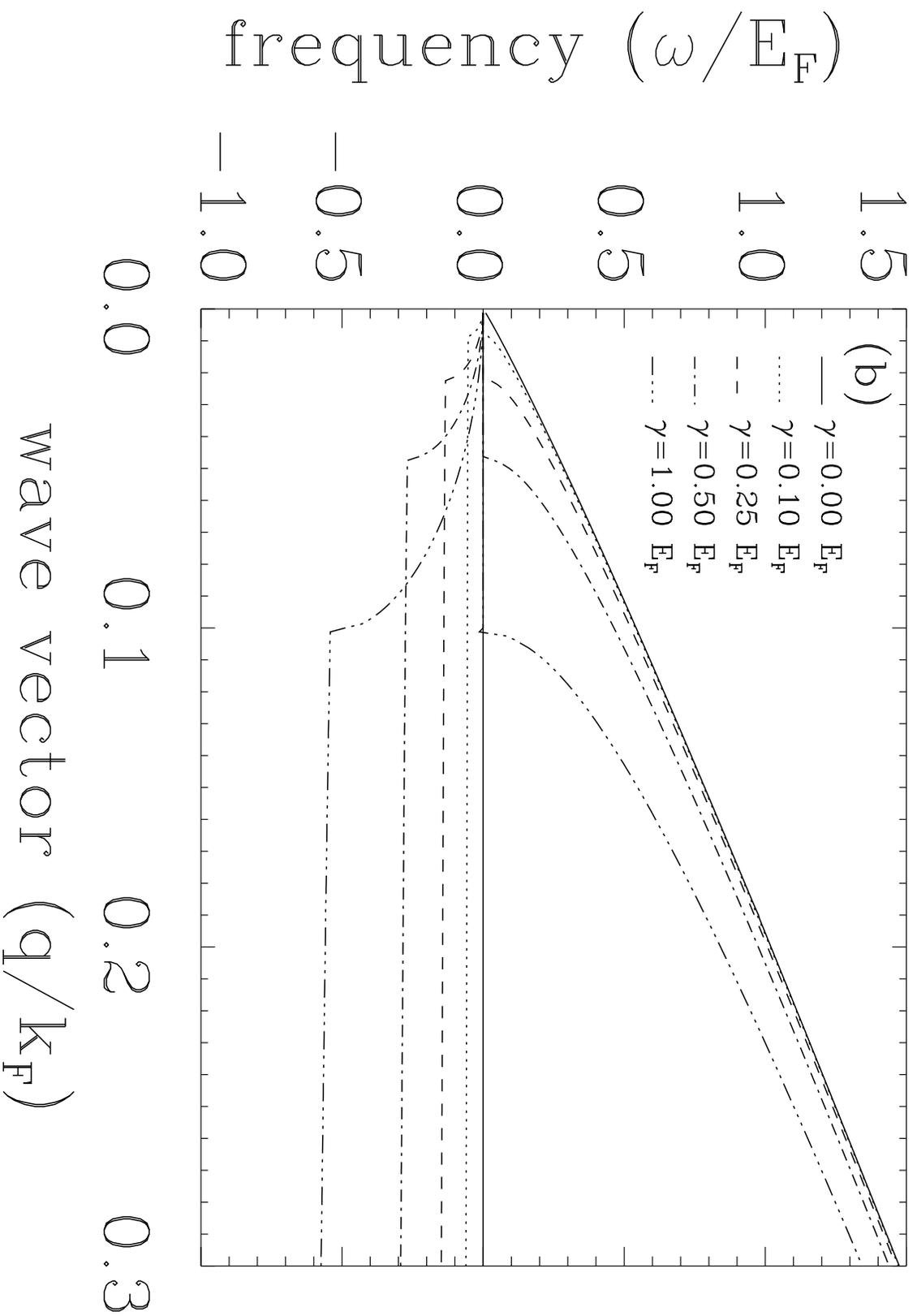
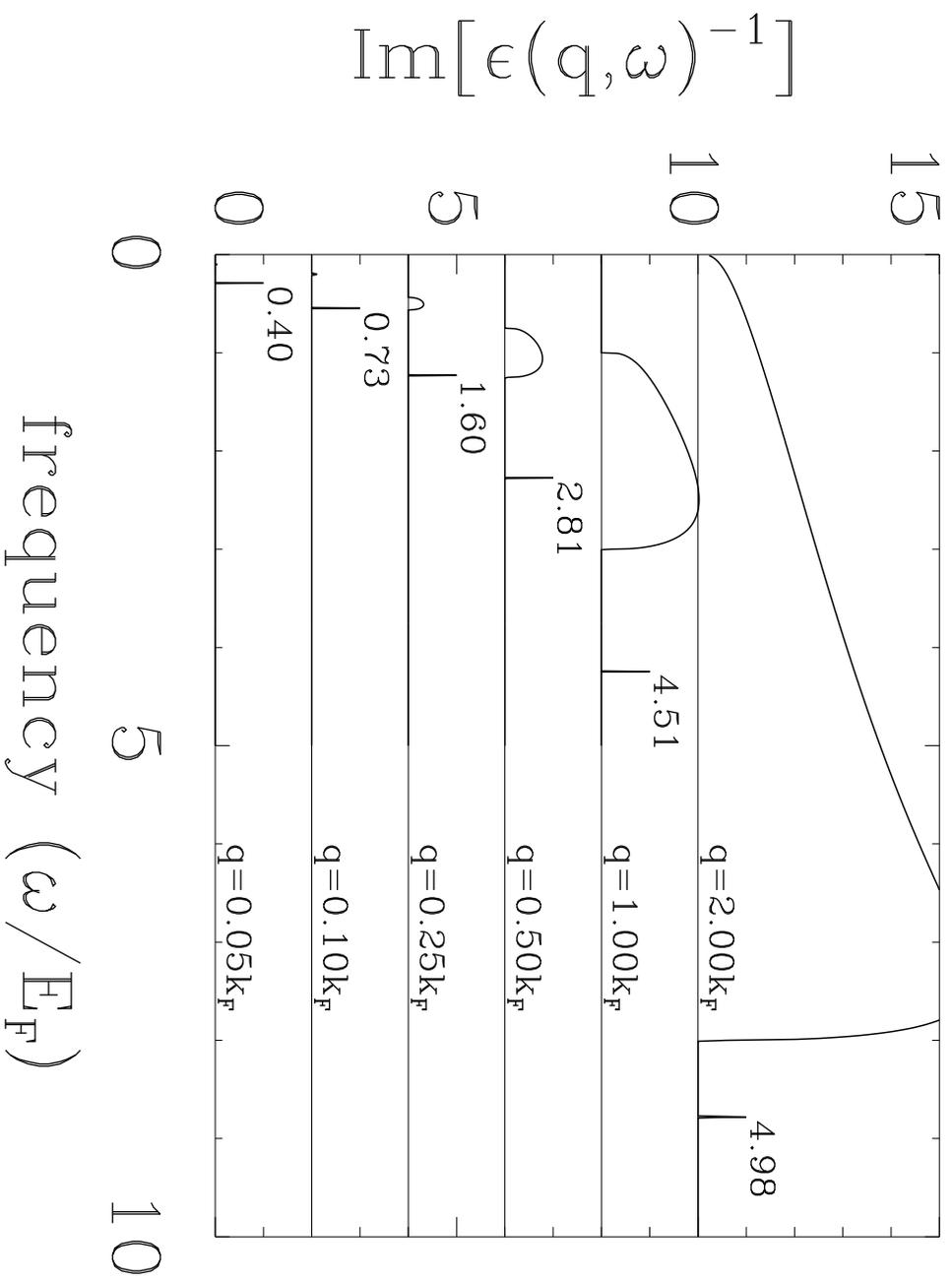
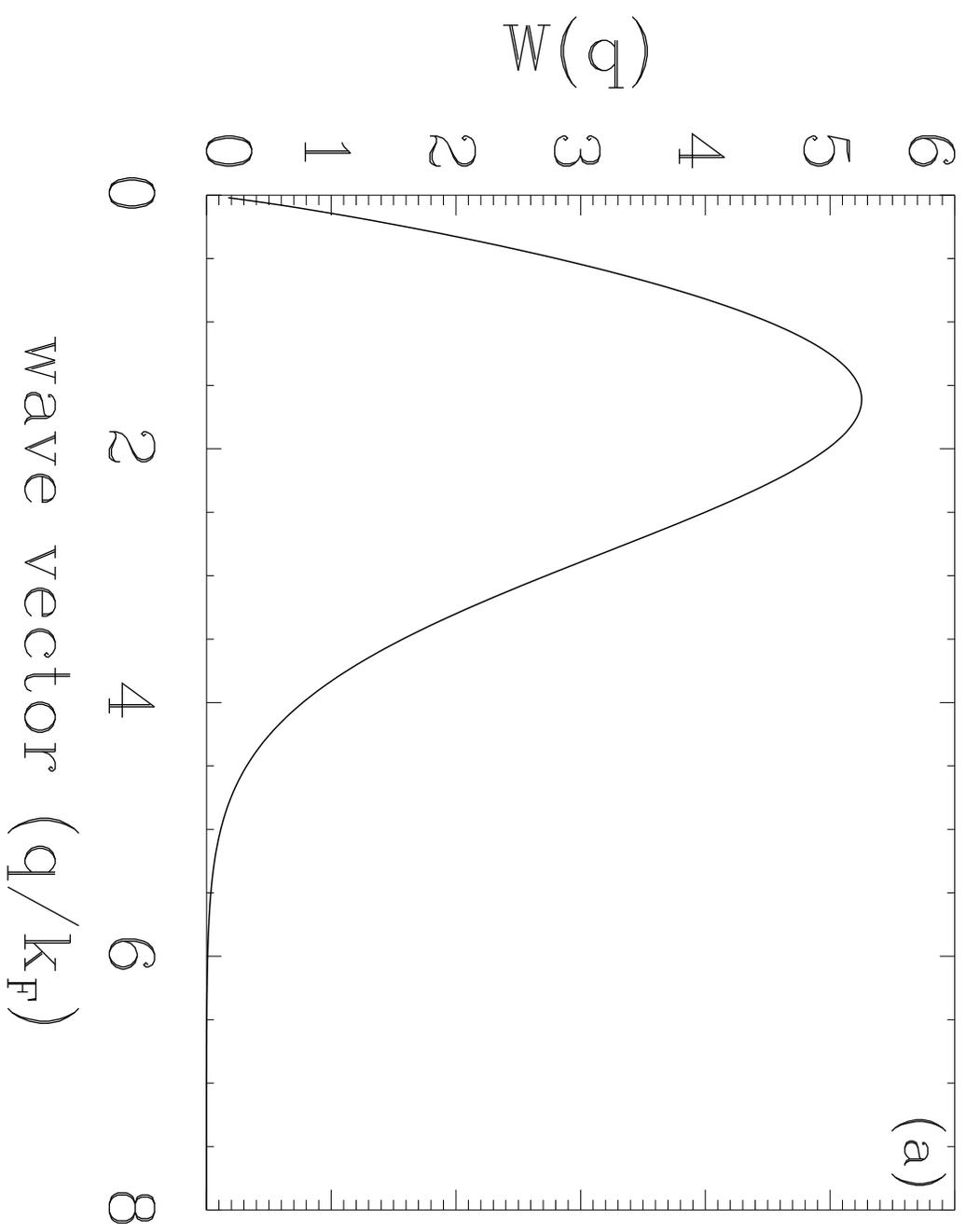
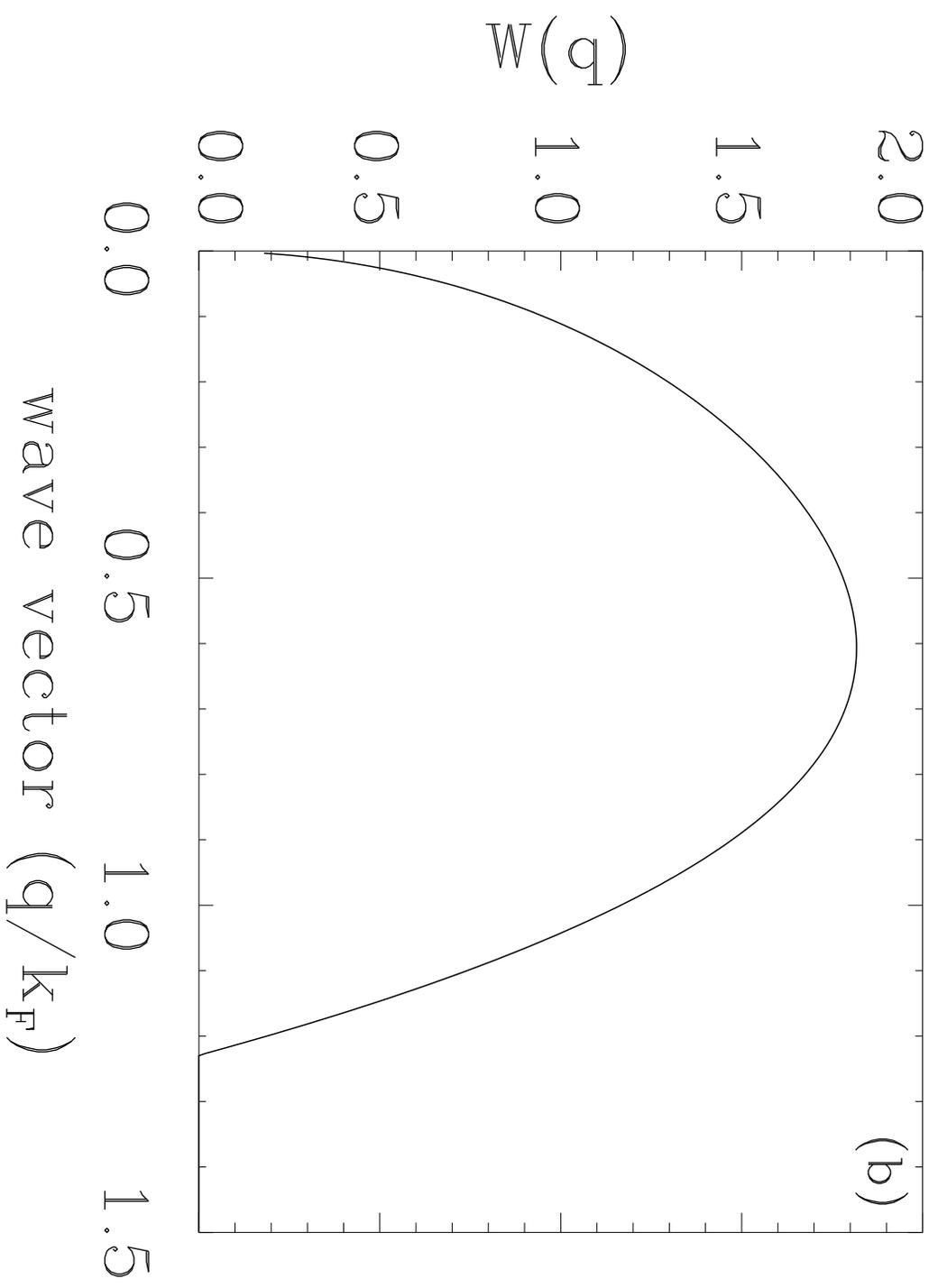
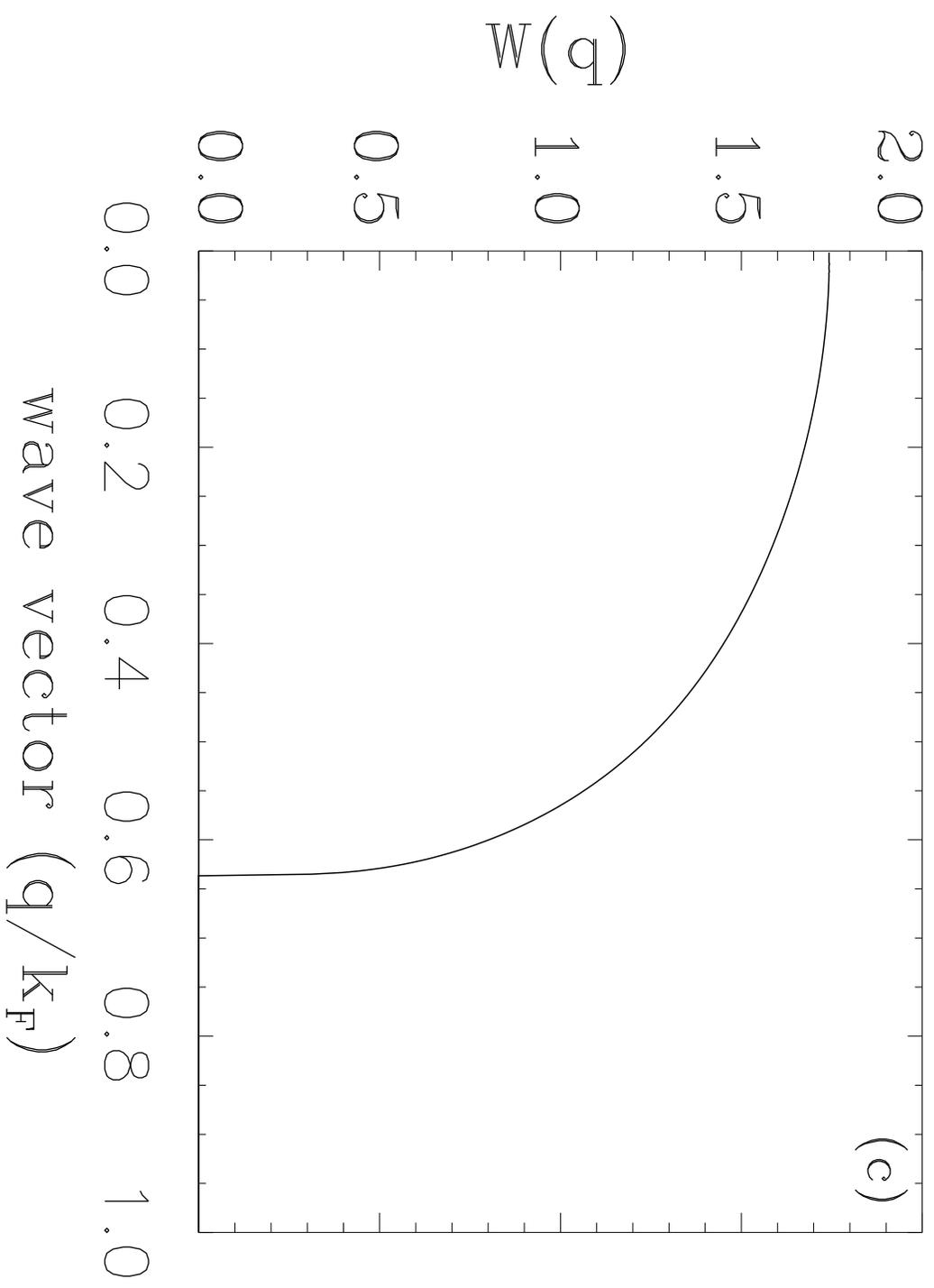
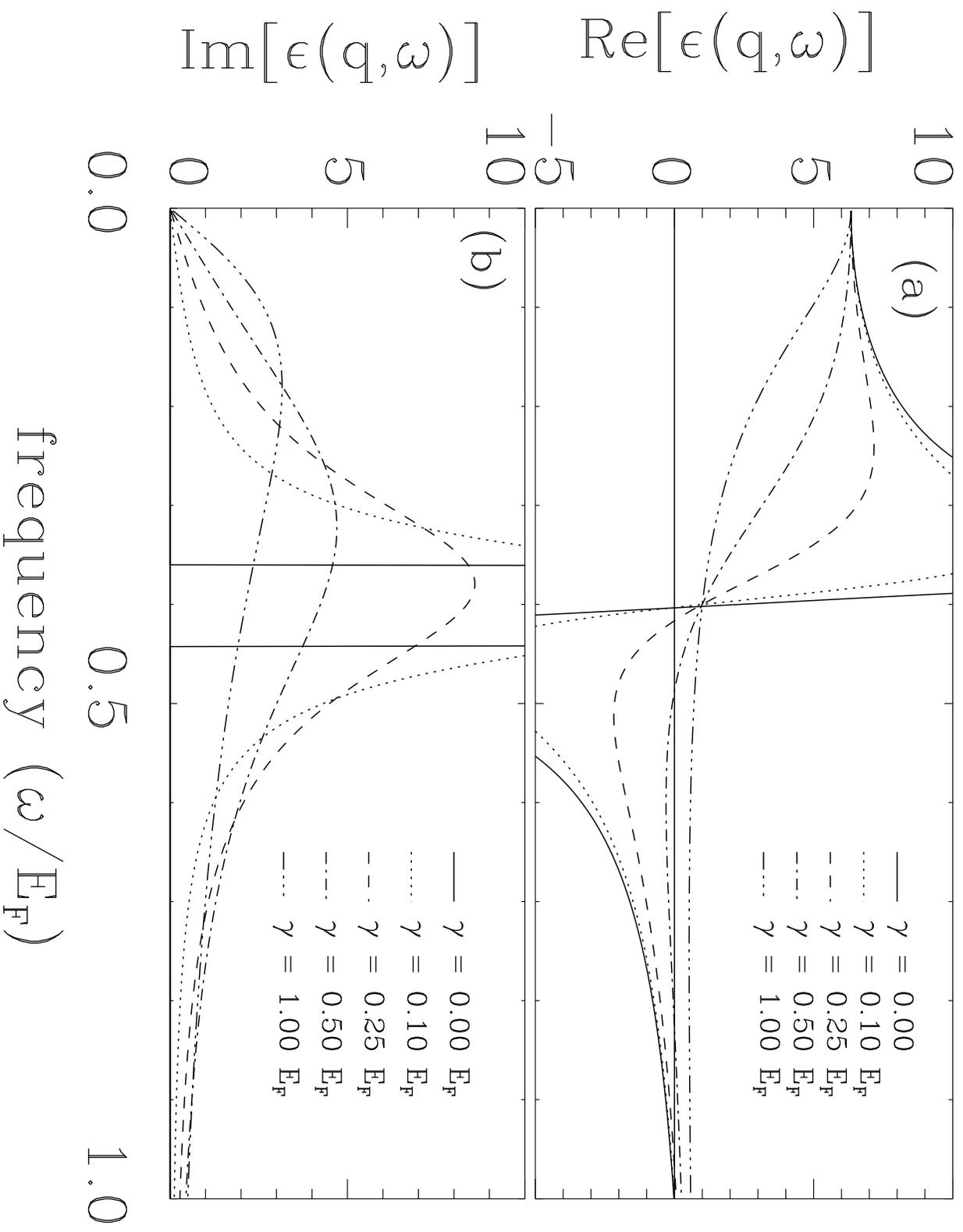
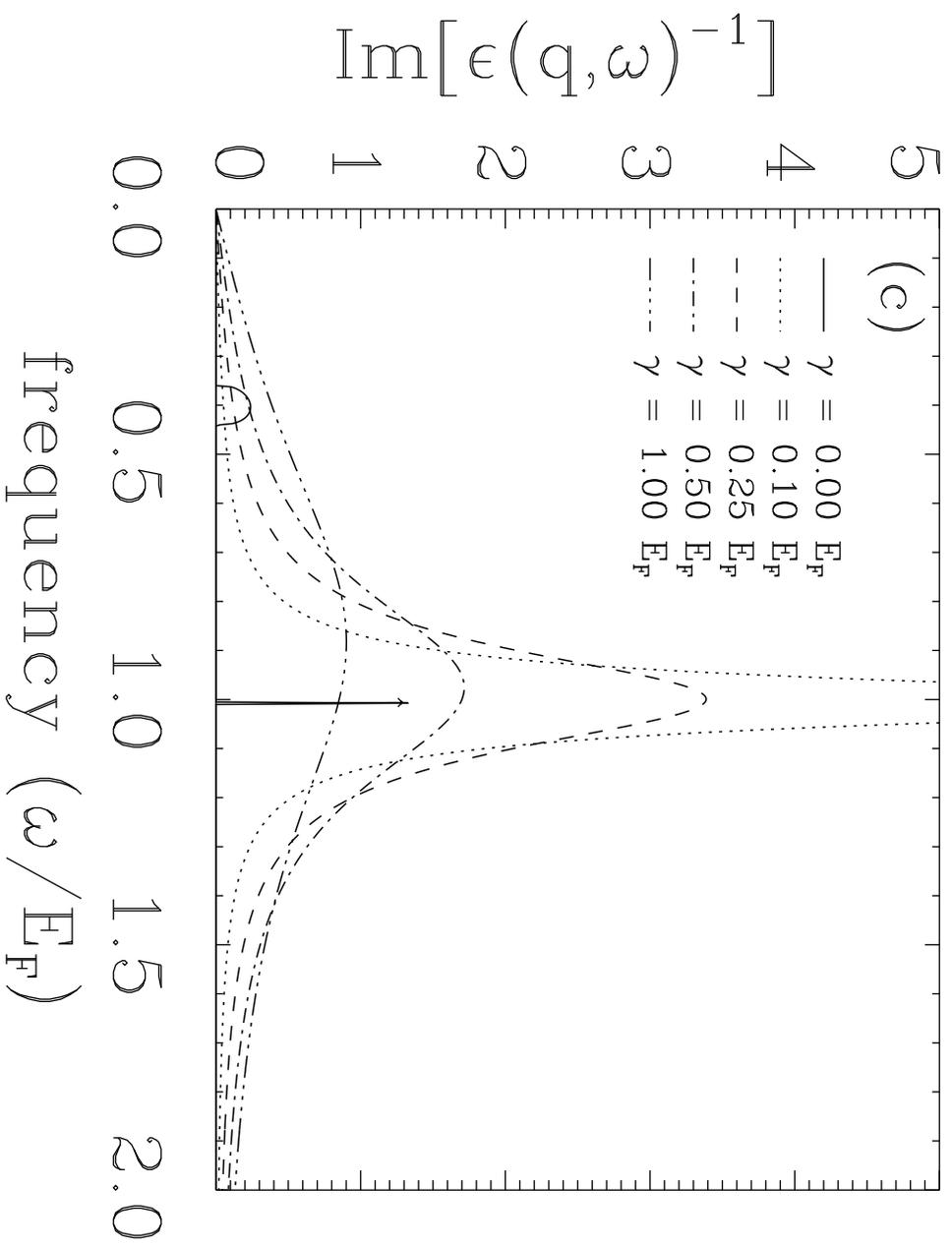
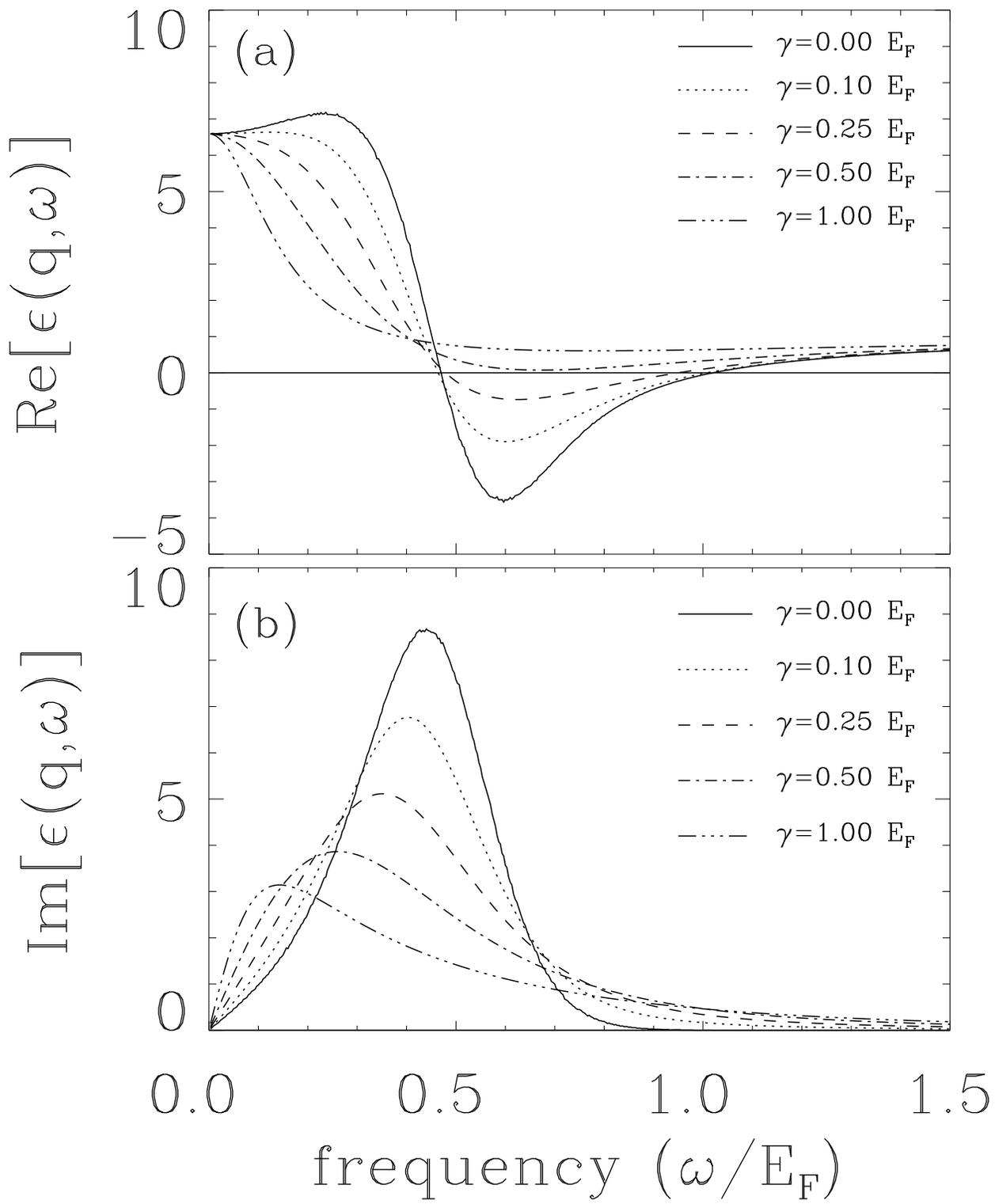
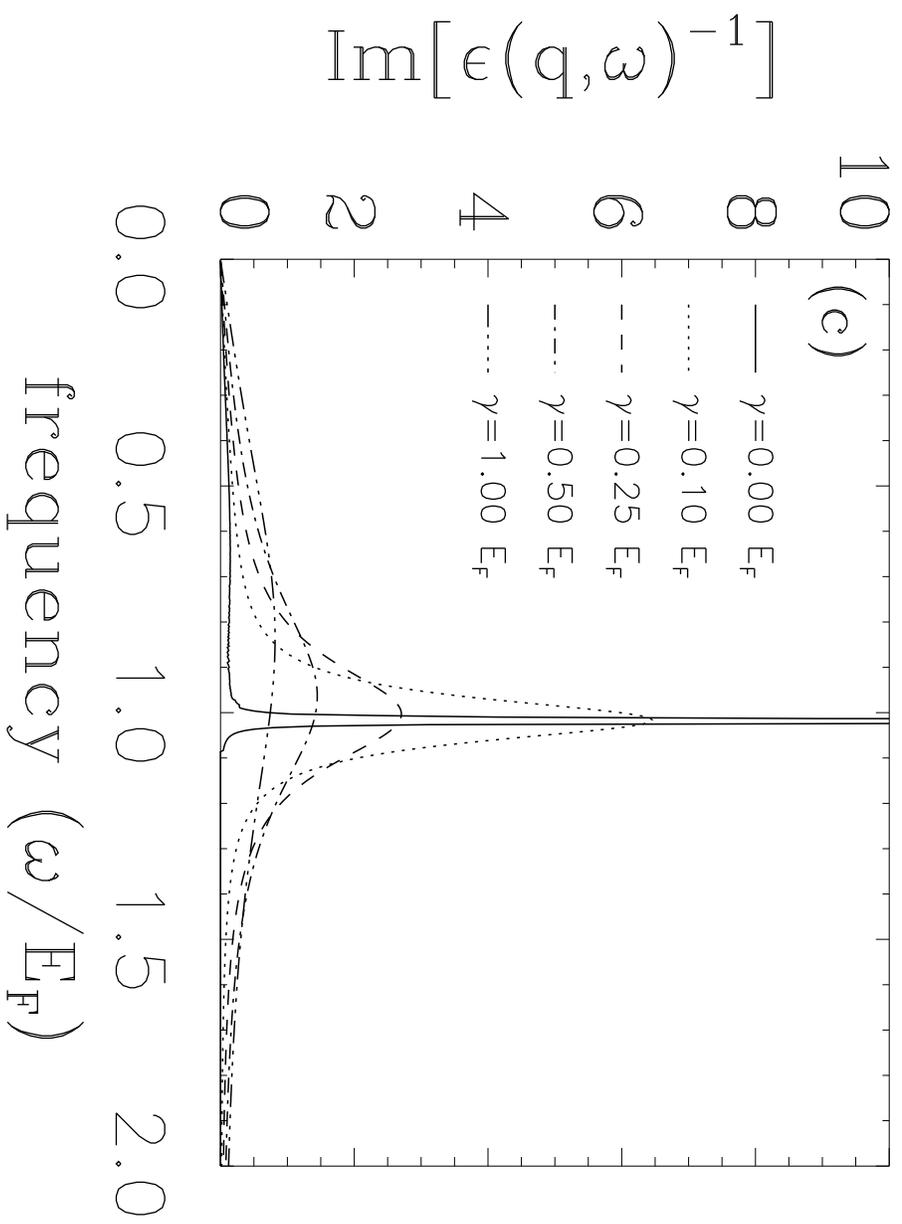
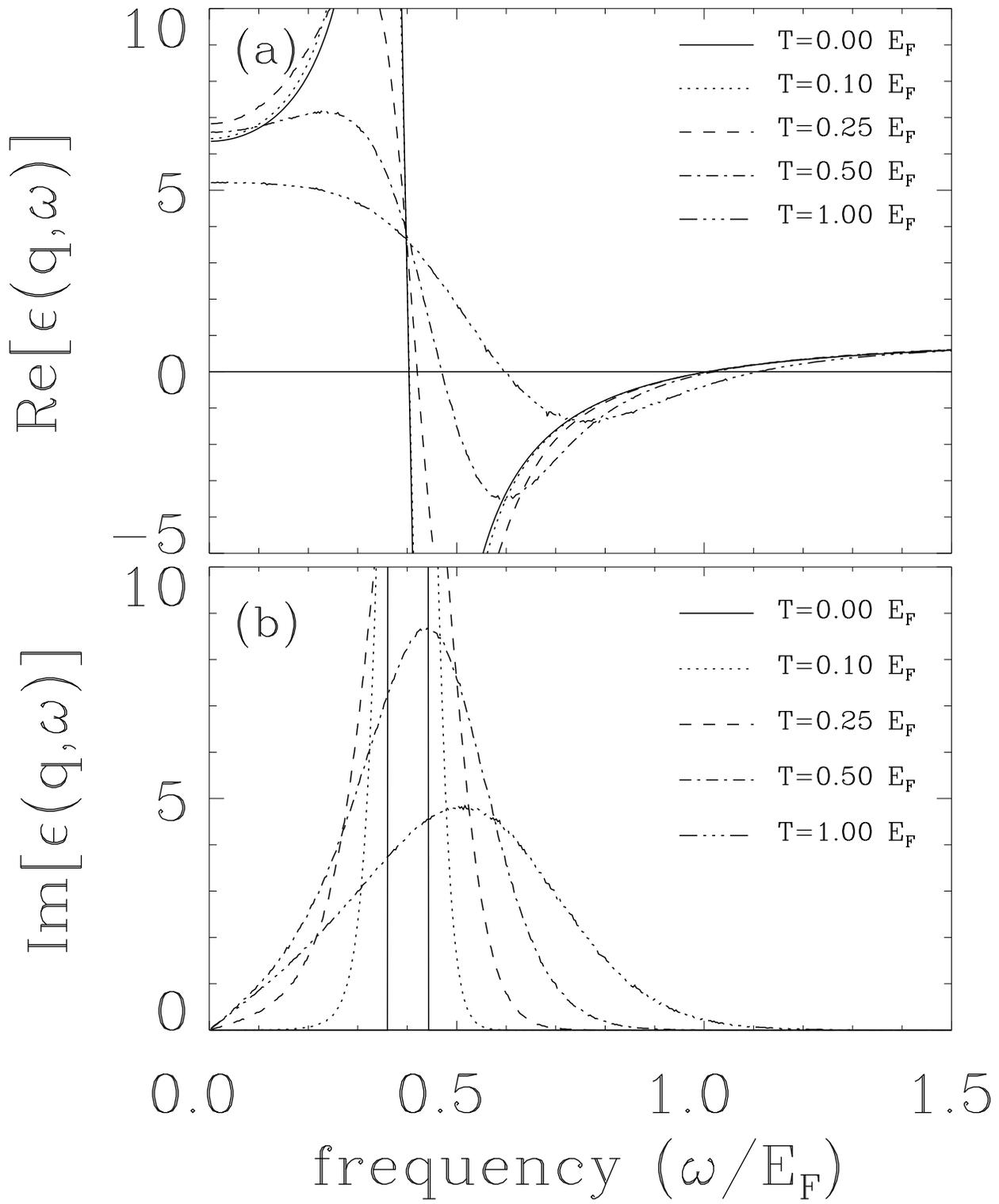
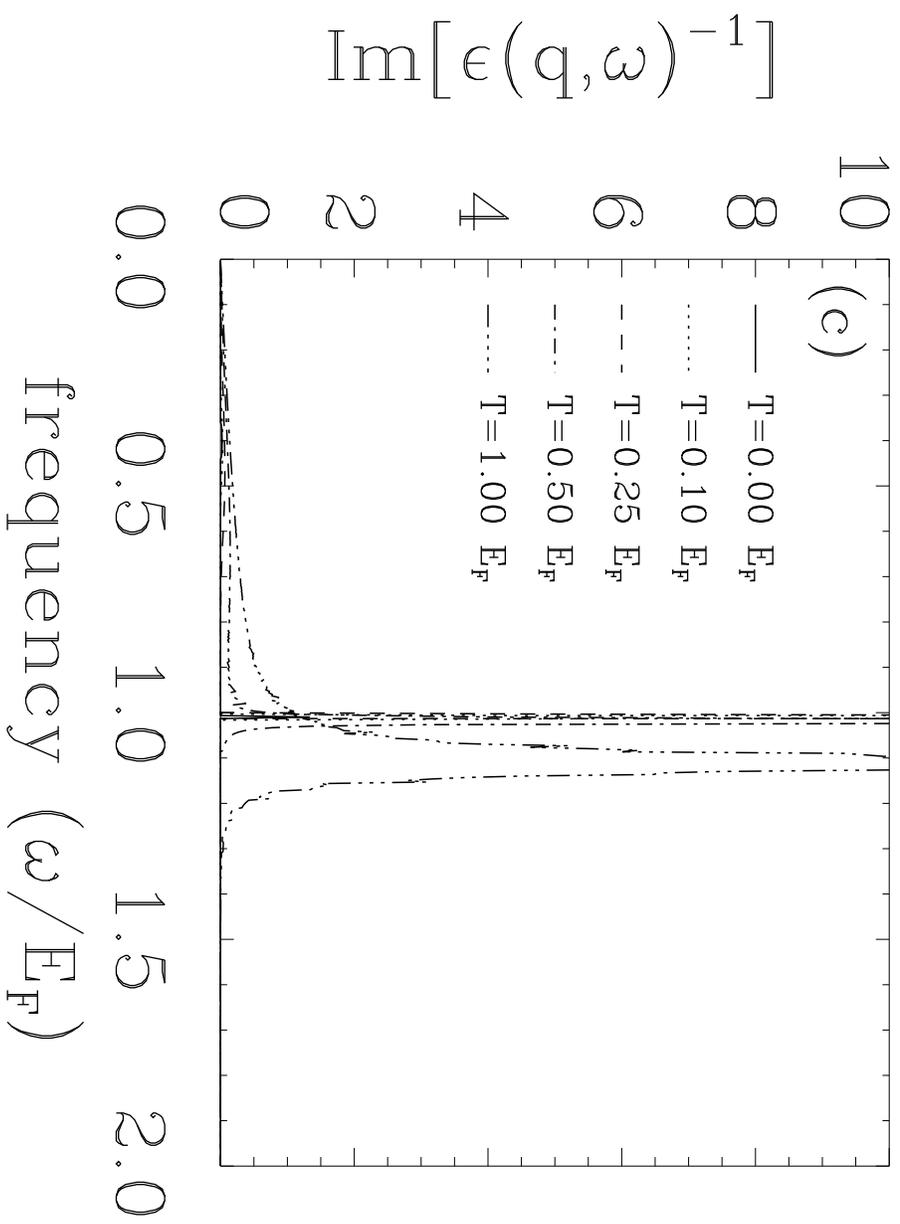
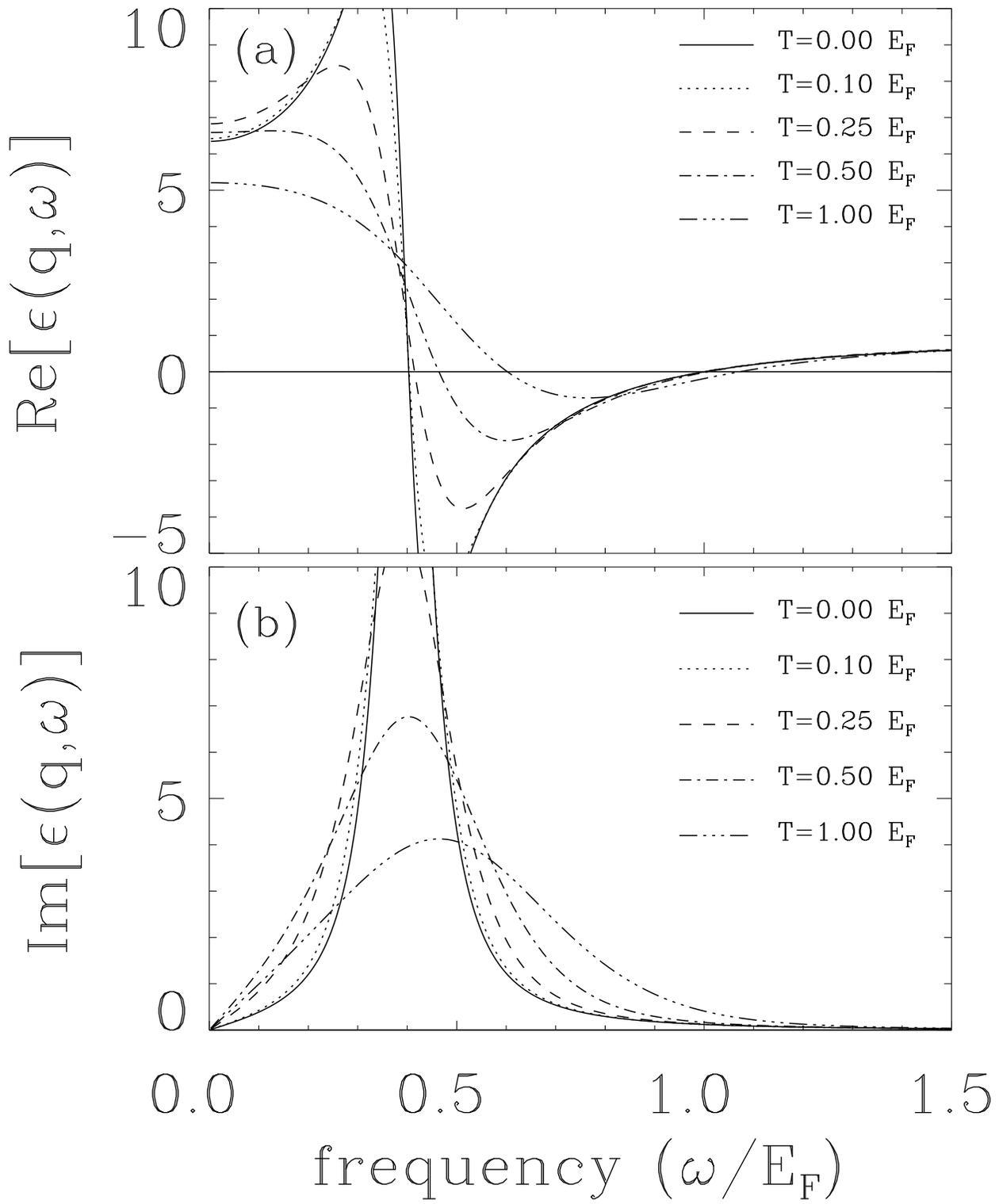
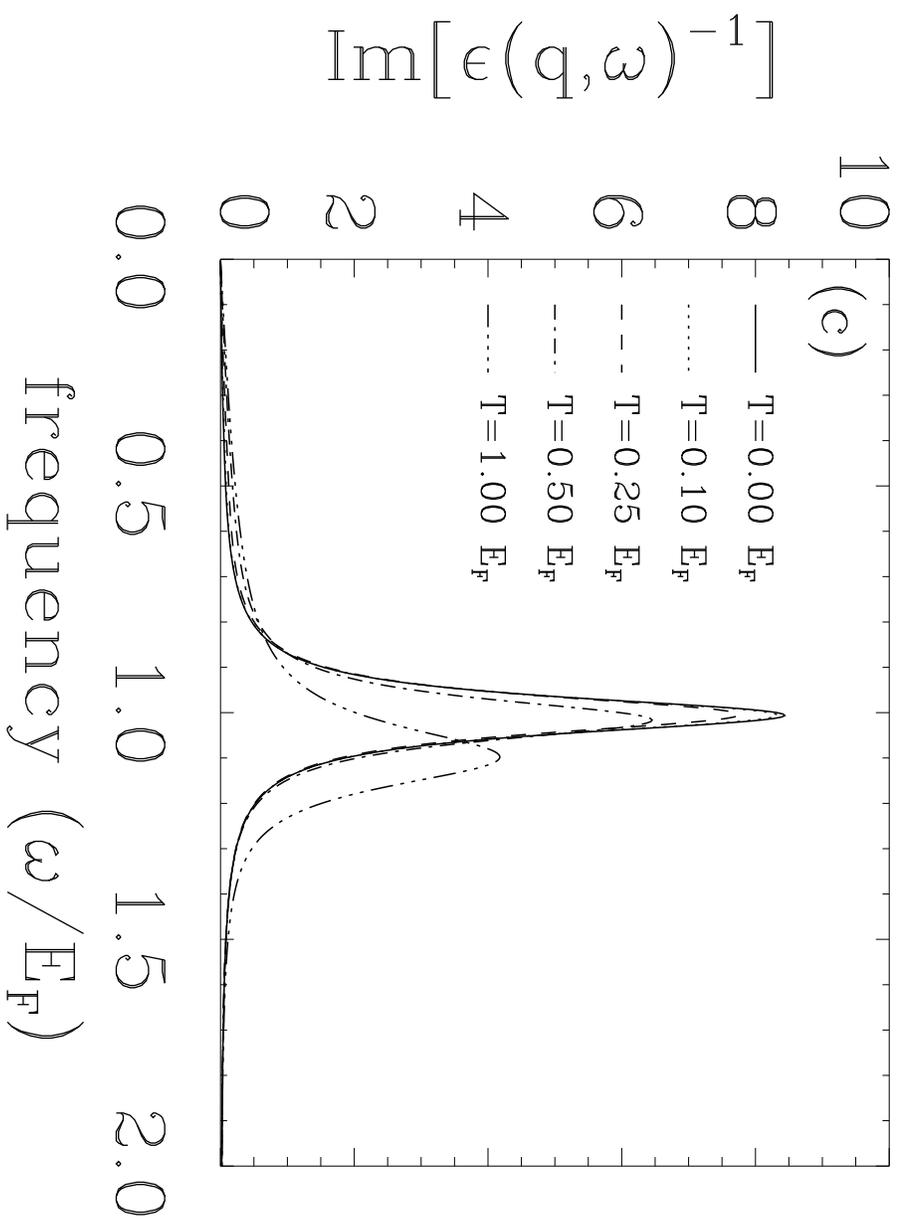
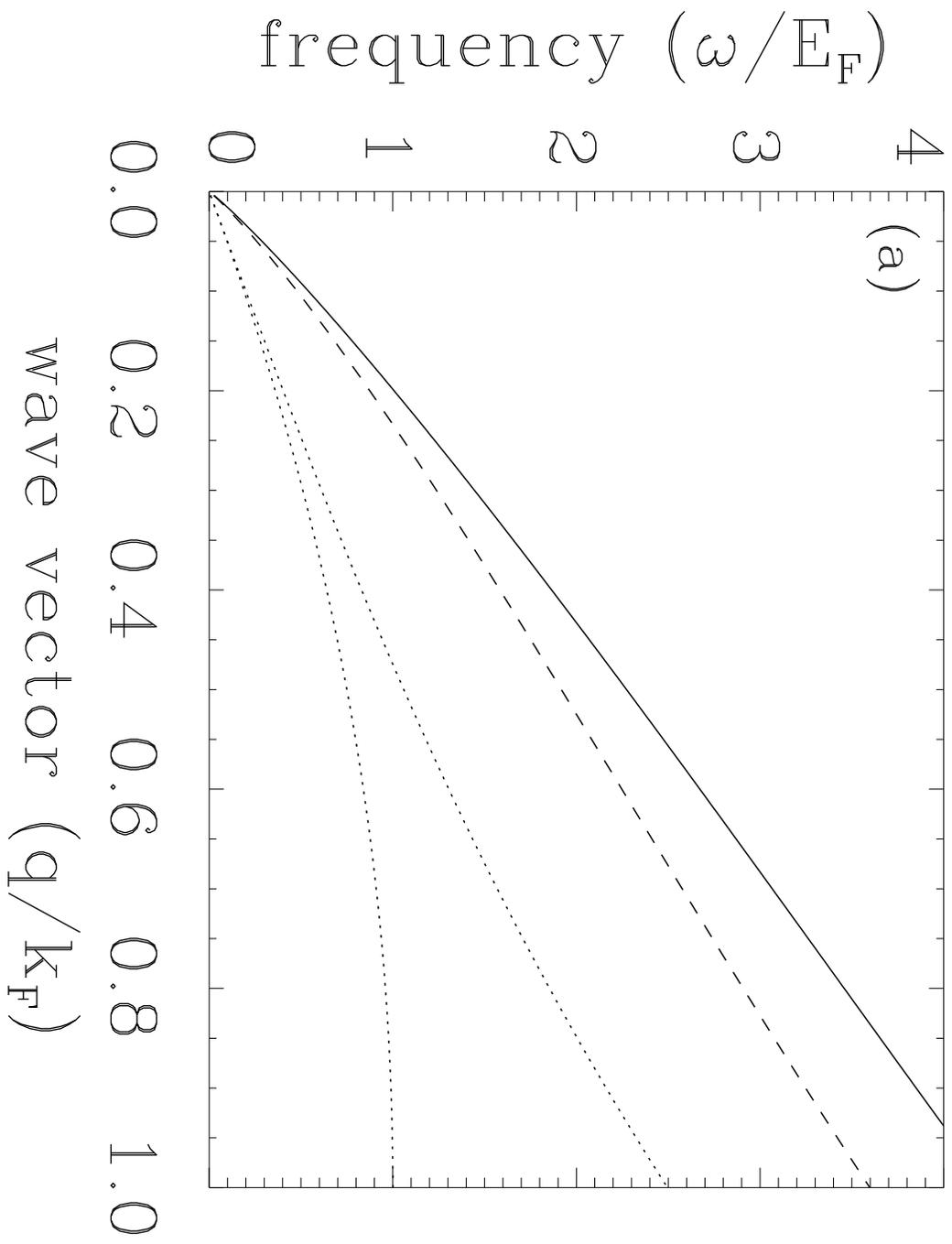
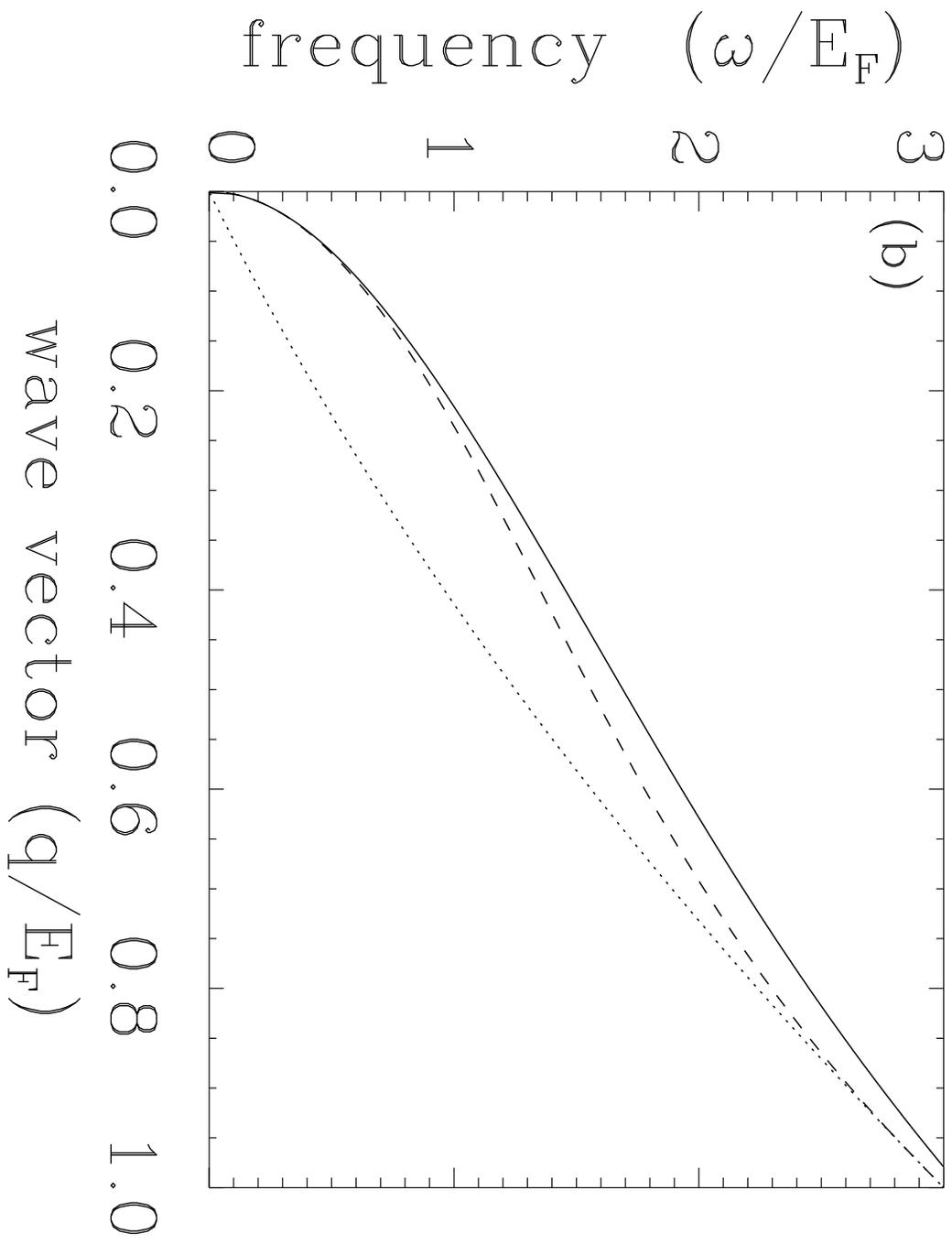
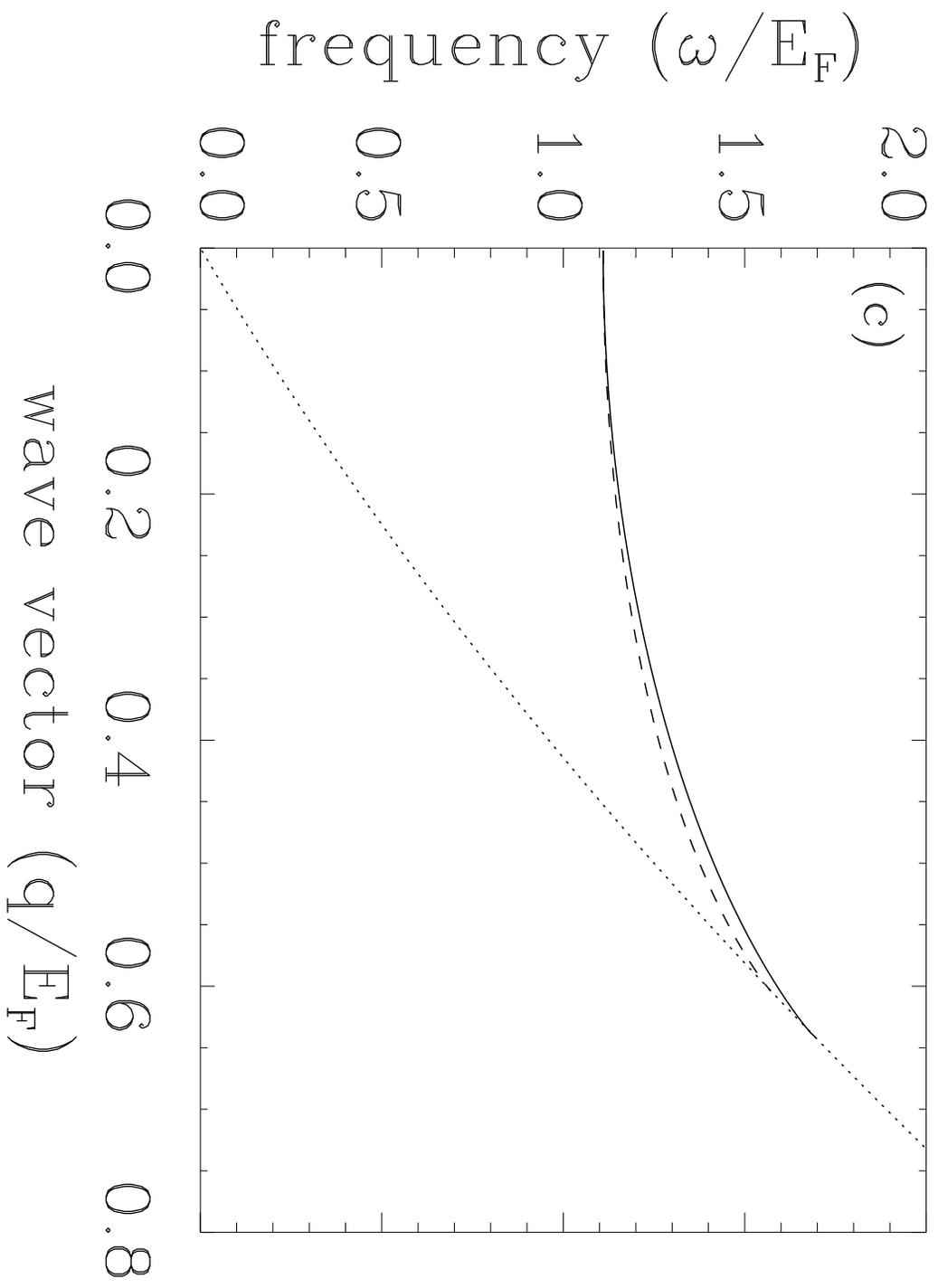
\end{document}